\documentclass[twocolumn]{aastex63}

\graphicspath{{./}{figs/}}

\usepackage{amsmath}
\usepackage{appendix}
\usepackage{needspace}

\newif\ifref
\reftrue
\reffalse
\definecolor{darkred}{rgb}{0.75, 0, 0}
\newcommand{\mb}[1]{\ifref\textcolor{darkred}{#1}\else #1\fi}

\newif\ifreff
\refftrue
\refffalse
\definecolor{darkred}{rgb}{0.75, 0, 0}
\newcommand{\mbb}[1]{\ifreff\textcolor{darkred}{#1}\else #1\fi}

\received{\today}

\submitjournal{the Astrophysical Journal Letters}

\shorttitle{The Potential of Asteroseismology to Resolve the Blue Supergiant Problem}
\shortauthors{Bellinger et al.}

\begin{document}
\title{The Potential of Asteroseismology to Resolve the Blue Supergiant Problem}

\correspondingauthor{E.\ P.\ Bellinger}
\email{earl.bellinger@yale.edu}

\author[0000-0003-4456-4863]{Earl Patrick Bellinger} 
\affil{Max Planck Institute for Astrophysics, Garching, Germany} 
\affil{Department of Astronomy, Yale University, USA} 
\affil{Stellar Astrophysics Centre, Aarhus, Denmark} 

\author[0000-0001-9336-2825]{Selma~E.~de~Mink}
\affil{Max Planck Institute for Astrophysics, Garching, Germany} 
\affil{Anton Pannekoek Institute for Astronomy and GRAPPA, University of Amsterdam, The Netherlands}

\author[0000-0002-8605-5285]{Walter E.~van~Rossem} 
\affil{Department of Physics \& Astronomy ``Augusto Righi,'' University of Bologna, Italy}
\affil{School of Physics and Astronomy, University of Birmingham, UK}
\affil{Stellar Astrophysics Centre, Aarhus, Denmark}

\author[0000-0001-7969-1569]{Stephen Justham}
\affil{Max Planck Institute for Astrophysics, Garching, Germany} 
\affil{School of Astronomy \& Space Science, University of the Chinese Academy of Sciences, Beijing, China}
\affil{National Astronomical Observatories, Chinese Academy of Sciences, Beijing, China}

\begin{abstract} 
Despite major progress in our understanding of massive stars, concerning discrepancies still remain between observations and theory. 
Most notable are the numerous stars observed beyond the theoretical main sequence, an evolutionary phase expected to be short lived and hence sparsely populated. 
This is the ``Blue Supergiant Problem.'' 
Stellar models with abnormal internal structures can provide long-lived solutions for this problem: core hydrogen-burning stars with oversized cores may explain the hotter ones, and core helium-burning stars with undersized cores may explain the cooler ones. 
Such stars may result from enhanced or suppressed mixing in single stars or, more likely, as the products of binary interaction and stellar mergers. 
Here we investigate the potential of asteroseismology to uncover the nature of blue supergiants. 
We construct stellar models for the above scenarios and show that they predict $g$-mode period spacings that differ by an order of magnitude: $\sim$200~min versus $\sim$20~min for long-lived core H and He burning stars, respectively. 
For the classical scenario of H-shell burning stars rapidly crossing the Hertzsprung gap, we furthermore predict changes of the order of $10^{-2}\,\mu\rm{Hz\,yr}^{-1}$ in high-frequency modes; this effect would be in principle observable from $\sim$5~yr of asteroseismic monitoring if these modes can be identified. 
This raises the possibility of revealing the internal structure of blue supergiants and thus determining whether these stars are indeed binary merger products. 
These asteroseismic diagnostics may be measurable through long time-series observations from the ongoing TESS mission and upcoming PLATO mission, thereby laying a path toward resolving the blue supergiant problem.
\end{abstract}


\section{Introduction} \label{sec:intro} 
Understanding the observed distribution of massive stars in the Hertzsprung--Russell diagram (HRD) still poses a major challenge, which is concerning for many fields that rely on predictions from massive stellar models \citep[see reviews by][]{2012ARA&A..50..107L, 2022ARA&A..60..455E}.  
This is particularly true for blue supergiants (BSGs), whose evolutionary status remains unclear. 

BSGs are among the brightest stars in the Universe, with temperatures of up to $25,000$~K and visible magnitudes of up to $M_{\rm V} = -9.5$~mag. 
They dominate the rest-frame visible light of distant star-forming galaxies and can be seen individually out to very large distances. 
The current record holder is Earendel, claimed to be a lensed BSG star at redshift $z=6$, observed by the Hubble and James Webb Space Telescopes \citep{2022ApJ...940L...1W}. 
In young nearby resolved stellar populations, BSGs are seen abundantly. They play a central role when inferring star formation histories \citep{2009AJ....138.1243H}, and are potentially valuable as distance indicators \citep{1999A&A...350..970K}. 
The strong winds of BSGs can give rise to bright X-ray sources when they are closely orbiting a neutron star or black hole \citep{2009MNRAS.398.2152D, 2017SSRv..212...59M}. 
They furthermore provide critical tests for stellar evolutionary models \citep{2019A&A...625A.132S, 2021A&A...650A.128G}. 

BSGs are responsible for some spectacular end points of stellar evolution. 
The most famous example is the progenitor star observed in pre-explosion images of SN~1987A, the nearest modern supernova to Earth \citep{1987ApJ...321L..41W}. 
The remnant is well known for showing several rings, and a binary-merger scenario provides a natural explanation for both that BSG star and the rings \citep{Podsiadlowski1992, Morris+PhP2007Science}.
Transient surveys show that a few percent of massive stars explode with very similar peculiar lightcurves \citep{2017ApJ...837..121G}, indicating that at least some massive stars end their lives as BSGs. 
More exotic is the suggestion that the collapsing core of a BSG may give rise to ultra-long gamma-ray bursts \citep{2018ApJ...859...48P}. 
BSGs have even been proposed as potential progenitors of the black holes with masses in the pair-instability gap that have been identified in gravitational-wave events \citep{2020ApJ...904L..13R, 2022MNRAS.516.1072C}. 

The classic theory of stellar evolution predicts that massive single stars become BSGs after exhausting their central hydrogen (H) supply. 
They fuse H in a shell surrounding an inert helium (He) core and expand across the Hertzsprung gap (HG). 
The problem is that this is predicted to be a very short-lived phase of stellar evolution \citep{1960MNRAS.120...22H, 1964ZA.....59..242H}. 
If BSGs are indeed these H-shell burning stars, then they should be $100-1000$ times rarer than MS stars, which clearly contradicts the fact that we observe them abundantly \citep{2014A&A...570L..13C, 2018ApJ...868...57C, 2023A&A...674A.212D}.  

Historically, most authors assumed BSGs to be massive stars in the long-lived central He-burning phase. 
This is indeed found in classic stellar models, which assume no or very little mixing beyond the convective core \citep{1987ApJ...321L..41W, 1989ApJ...339..365W, 2019A&A...625A.132S}. 
A well-known problem with these models is that they fail to reproduce the width of the main sequence \citep[MS; e.g.][]{2020MNRAS.496.1967K, 2021A&A...648A.126M}. 
Moreover, they predict a gap between the end of the MS and the location where He-burning stars reside \citep{1991A&A...252..669L}, which is not observed \citep{1990ApJ...363..119F}. 

In recent years, evidence has been mounting in favor of extra mixing. 
This increases the core masses and extends the end of the MS to cooler temperatures, as well as to larger luminosities and radii \mbb{\citep{2023NatAs...7..913B}}. 
These models can explain the hotter BSGs \citep{2020MNRAS.496.1967K, 2011A&A...530A.115B}, but it remains challenging to explain the very gradual decrease in numbers towards cooler temperatures. 

The above solutions to the BSG problem are based on models of isolated stars. 
We know, however, that the majority of young massive stars are found in close binaries, which will interact and may even merge \citep{1992ApJ...391..246P, 2007ApJ...670..747K, 2012Sci...337..444S}. 
Understanding the distribution of stars in the HRD thus requires consideration of the effect of binarity. 
For example, binary interactions may lead to MS stars that experience extra mixing \citep{2007A&A...465L..29C, 2009A&A...497..243D, 2013MNRAS.434.3497G, 2013ApJ...764..166D}. 
The resulting widening of the MS may be responsible for the hotter BSGs. 

Interactions of post-MS stars may also produce BSGs via mass transfer or merging \citep{1992ApJ...391..246P, 1995A&A...297..483B, 2013A&A...552A.105V, 2014ApJ...796..121J}. 
Mergers which occur after the MS and before the end of core-He fusion are called Case~B mergers.  
Stars resulting from such mergers tend to remain compact during the long-lived phase of central He burning if the resulting core-to-envelope mass ratios are sufficiently small \citep{Giannone+1968, Lauterborn+1971, 2019A&A...621A..22F}. 
The B[e] supergiant in the Small Magellanic Cloud system R4 may be the outcome of such a merger \citep{Langer+Heger1998,Pasquali+2000}. 
Such mergers are promising candidates to solve the BSG problem, especially at cooler temperatures. 

In this work, we conduct a proof-of-principle study to explore the potential of asteroseismic constraints to distinguish various scenarios that have been proposed to explain BSGs. 
We construct grids of stellar evolution models for each of these scenarios, and perform stellar oscillation calculations on the models. 
Ultimately, we conclude that asteroseismic analysis of BSGs has the potential to distinguish these scenarios, and hence to resolve the blue supergiant problem.

\section{Evolutionary Models} \label{sec:models} 
\begin{figure*}
    \centering%
    \includegraphics[height=10cm, trim={0 0 2.8cm 0}, clip, keepaspectratio]{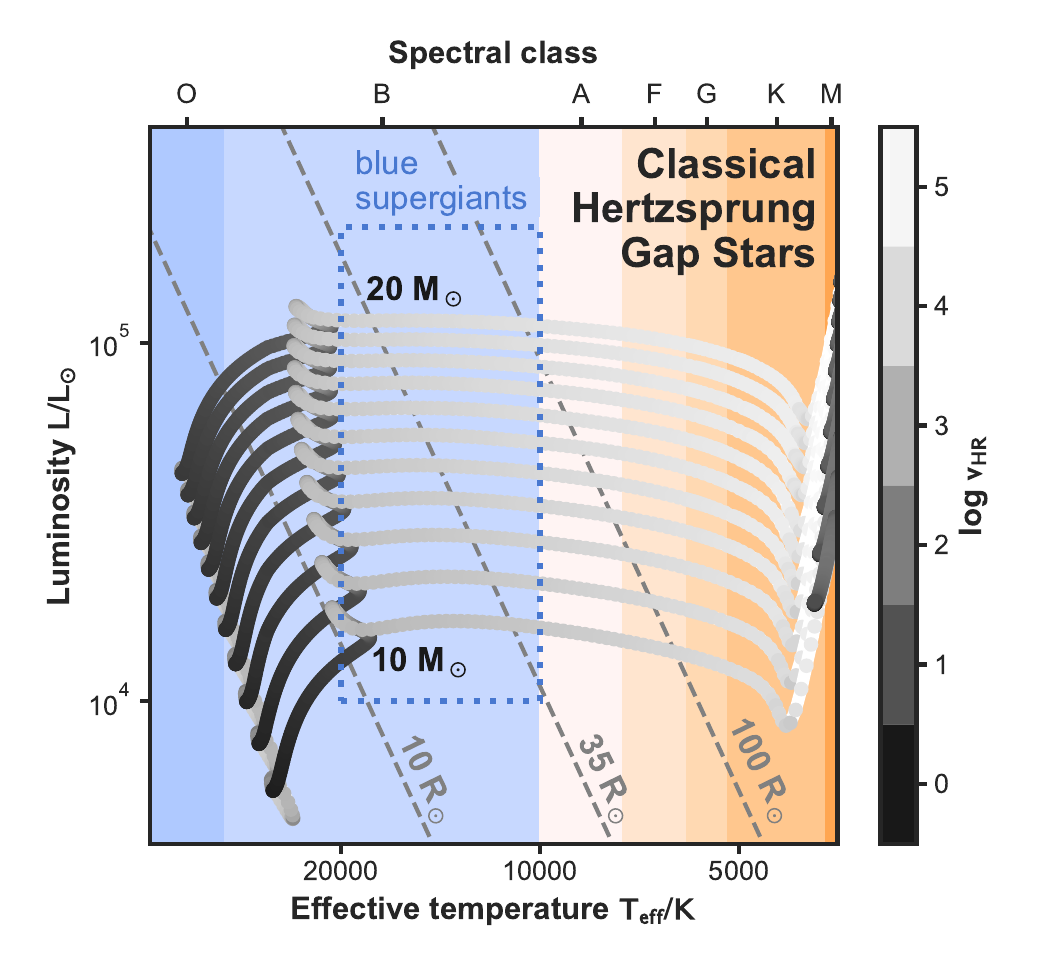}%
    \includegraphics[height=10cm, trim={2.6cm 0 0 0}, clip, keepaspectratio]{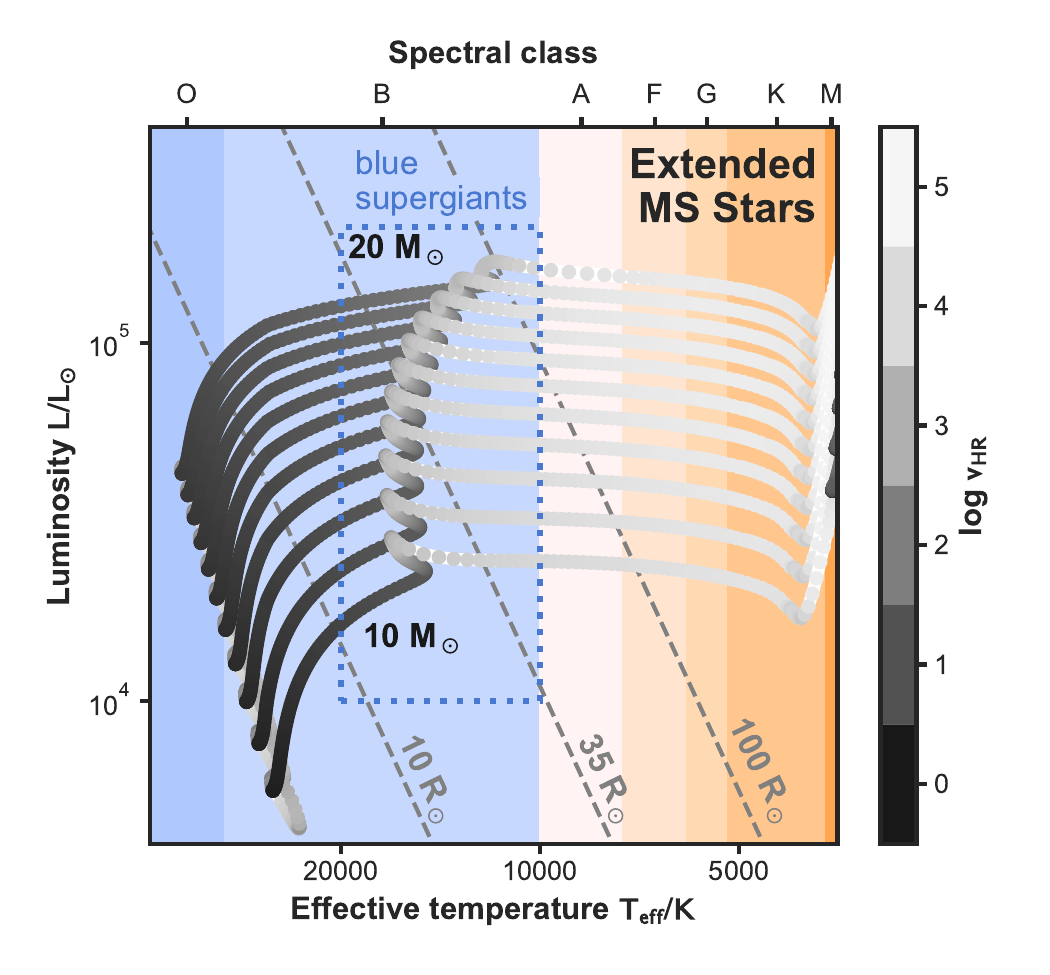}\\%
    \includegraphics[height=10cm, trim={0 0 2.8cm 0}, clip, keepaspectratio]{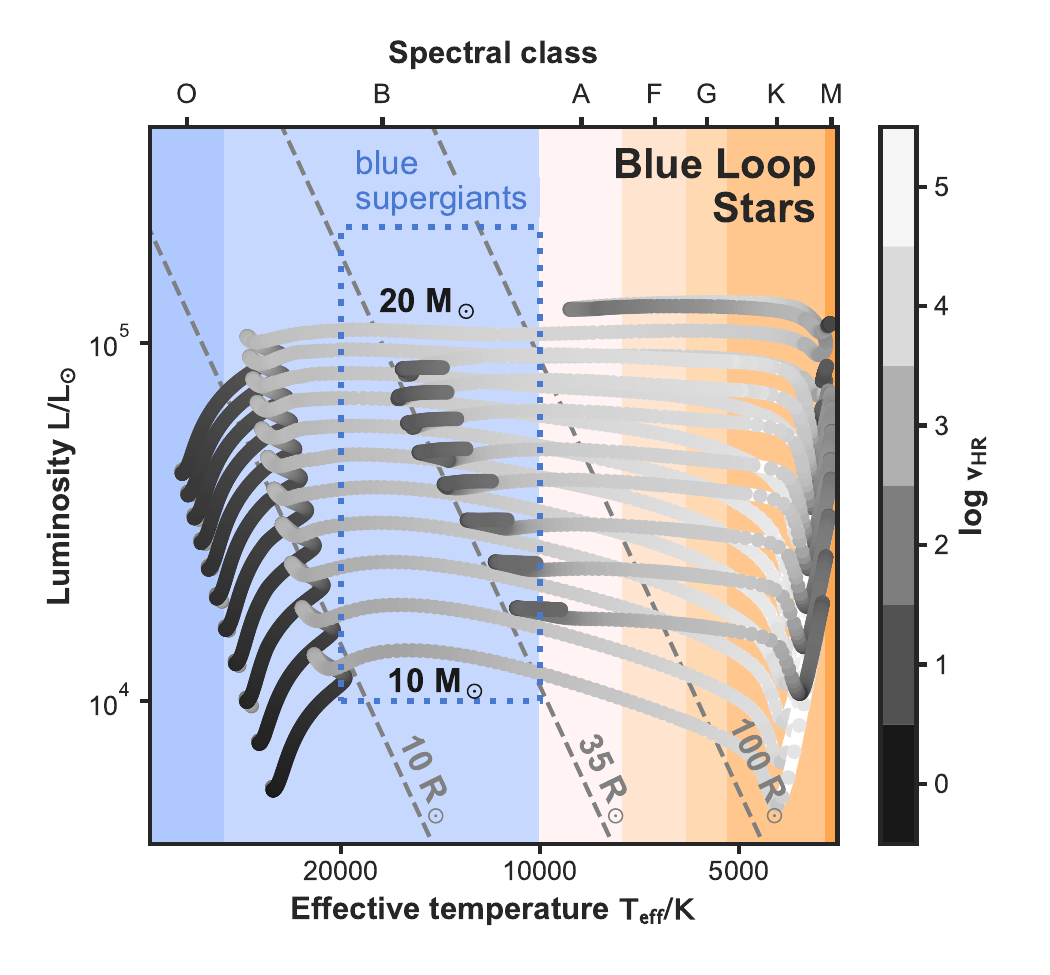}%
    \includegraphics[height=10cm, trim={2.6cm 0 0.2cm 0}, clip, keepaspectratio]{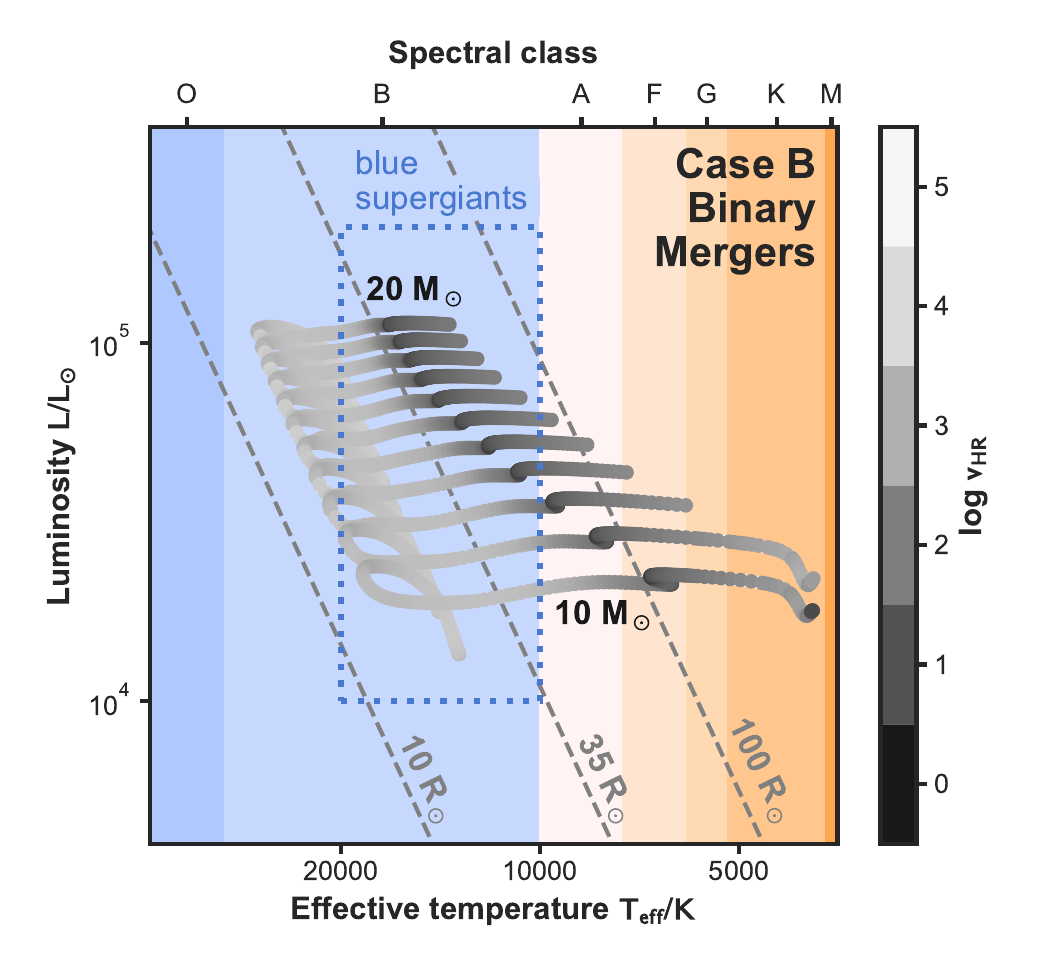}%
    \caption{%
    Hertzsprung--Russell diagrams for scenarios resulting in blue supergiants: classical Hertzsprung gap stars (top left panel), extended main-sequence stars (top right panel), blue loop stars (bottom left panel) and Case~B binary mergers (bottom right panel). 
    The tracks are of $10-20$~M$_\odot$ and are computed until core-He exhaustion. 
    The tracks are colored by their ``velocity'' in the HRD (Equation~\ref{eq:hrv}), which is a proxy for time spent and hence number of expected stars observed (darker means longer-lived and more stars). 
    The higher-mass tracks are the more luminous ones, and evolution generally proceeds from hot to cool. 
    The blue supergiant problem is an observed overdensity of stars in the Hertzsprung gap ($\sim$10,000~--~20,000~K as indicated by the dotted blue box), as isolated stars with normal cores (top left panel) should not numerously populate this region. 
    \label{fig:hrd} }%
\end{figure*}

Here we describe the construction of the stellar evolutionary models for the four pathways that we have described to blue supergiants. 
The first are classical Hertzsprung gap stars: stars that have exhausted their core-H supply and cross the HG as H-shell burning stars. 
The second are main-sequence stars with extended convective cores; the third are post-MS stars in a blue loop phase; and the fourth are Case~B binary mergers (see Figure~\ref{fig:hrd}). 

\subsection{Isolated stars}
We calculate evolutionary tracks for stars with masses ranging from 10 to 20~M$_\odot$. 
We use Modules for Experiments in Stellar Astrophysics \citep[\textsc{Mesa} r23.05.1,][]{2011ApJS..192....3P, 2013ApJS..208....4P, 2015ApJS..220...15P, 2018ApJS..234...34P, 2023ApJS..265...15J} to simulate their evolution. 
We consider here models of solar metallicity because the most promising targets for asteroseismic characterization generally belong to the solar neighborhood. 
We treat convection using the mixing-length theory \citep{1953ZA.....32..135V} \mb{with a mixing length of 2 pressure scale heights}. We include \mb{radiative step} overshooting of convection zones, for which we use a value of 0.335 pressure scale heights \citep{2011A&A...530A.115B} for the tracks of classical Hertzsprung gap stars. We use a larger factor of 0.7 for the extended main-sequence tracks, and a smaller factor of 0.1 for the blue loop tracks. 
We use the Vink \citep{2001A&A...369..574V} prescription for mass loss with a scaling factor of 1. 
We include the effects of thermohaline mixing \citep{1980A&A....91..175K} with its associated efficiency parameter set to 2. 
We include the effects of semi-convection \citep{1985A&A...145..179L} with its efficiency set by 1 for all except for the blue loop tracks, for which we use a value of 100 \citep{2019A&A...625A.132S}. 
\mb{We neglect the important yet poorly-understood effects of rotation in our models, as the transport of angular momentum in stellar interiors remains an unsolved problem \citep[][see also Section~\ref{sec:discussion} below for further discussion]{2019ARA&A..57...35A}.} 
We halt the evolution at core-He exhaustion. 
\mb{Timestepping in the models is adjusted adaptively; we assessed the temporal convergence of our models by modifying the time resolution and verifying that the conclusions drawn from the models are the same.} 

\subsection{Merger models} \label{sec:merger}
Case B merger events occur when the primary star in a binary system has exhausted its central hydrogen supply, leading to envelope expansion. As the primary star swells, it may fill its Roche lobe and initiate mass transfer to its less evolved secondary companion, which may lead to the stars merging \citep[see, e.g.,][and references therein]{2014ApJ...796..121J}. 

Here we explore merger products with resulting masses in the same range as in the isolated star case, i.e., $10$~--~$20$~M$_\odot$. 
We do not model the merging process itself; instead, we model the structure of the resulting object based off 3D hydrodynamic simulations of stellar mergers \citep[e.g.,][see also \citealt{2013MNRAS.434.3497G, 2023ApJ...942L..32R} for similar approaches]{2019Natur.574..211S}. 
We then use these structures as the initial models for our stellar evolution simulations in order to study the subsequent post-merger evolution and asteroseismology. 

\begin{figure*}
    \centering
    \includegraphics[width=\linewidth]{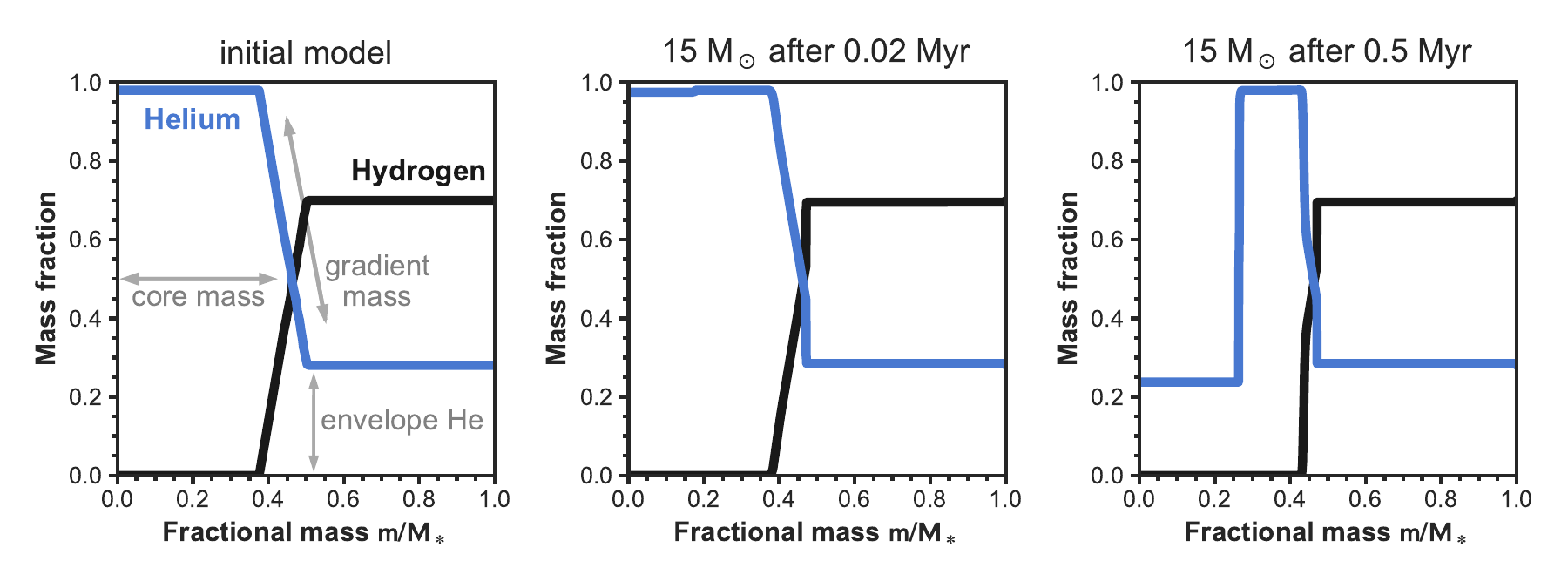}
    \caption{Anatomy of our schematic merger models. 
    \textsc{Left panel.} The initial chemical profile consists of a He core (of variable mass) below an outer envelope (of variable He abundance) separated by a gradient (also of variable mass). \textsc{Middle panel.} Internal structure of the model shortly after thermal relaxation.  \textsc{Right panel.} The state after 0.5~Myr. Most of the center in the core has undergone fusion. Notice that despite an initial gradient, the system evolves to a more step-like profile. } 
    \label{fig:merger}
\end{figure*}

We parameterize the merger product by a step- and slope-like profile consisting of a helium core and a hydrogen envelope, each over a background of solar metallicity. 
A schematic of the initial merger model structure is given in Figure~\ref{fig:merger}. 
We consider variations in the initial ratio of the He core mass to the total stellar mass $M_c/M_\ast$ in the range of $0.2$~--~$0.4$. 
We vary the fraction of helium $Y_{\rm{s}}$ in the outer envelope from $0.28$ to $0.34$, which represents mixing of helium into the envelope. 
We also consider models with a gradient between the helium core and the hydrogen envelope, with the mass of the region over which we impose that gradient, $M_g$, ranging from $0$ to $0.3$~$M_\ast$. 
The physics considered in the simulations of the subsequent evolution are the same as in the isolated star case described above.

Our code and models are publicly available\footnote{\url{https://github.com/earlbellinger/blue-supergiants} \mb{\citep{bsg-zenodo}}}.

\subsection{HRD Position \& Timescales}
In order to estimate how many stars we expect to observe in a particular region of the Hertzsprung--Russell diagram, we define a ``velocity'' in the HRD as 
\begin{equation} \label{eq:hrv}
    v_{\rm{HR}} 
    =
    \dfrac{1}{C\Delta t}
    \sqrt{
        \left[
            \Delta\log \left(L/\rm{L}_\odot \right)
        \right]^2
        +
        \left[
            \Delta\log\left(T_{\rm{eff}}/\rm{K} \right)
        \right]^2
    }
\end{equation}
where $C$ is a normalization factor set to be the zero-age main sequence (ZAMS) velocity of a $10~\rm{M}_\odot$ single-star track \citep{Walter, 2022arXiv220911502G}. Hence, for a particular region in the HRD and for the same constant birthrate, a value of for example ${\log v_{\rm{HR}} = 0}$ means that we expect to see as many stars in that position as we do for a $10~\rm{M}_\odot$ star at ZAMS. On the other hand, a value of 3 means that we expect to see a thousand times fewer stars in that position. 
We may consider a phase long-lived when $\log v_{\rm{HR}} \lesssim 2$. 

Figure~\ref{fig:hrd} compares evolutionary tracks of the four scenarios resulting in BSGs in a Hertzsprung--Russell diagram. 
The merger models shown in this figure have ${M_c/M_\ast = 0.3}$, ${Y_{\rm{s}} = 0.28}$, and ${M_g=0}$. 
Here it is evident that classical Hertzsprung gap evolutionary tracks proceed quickly through the BSG phase before undergoing He fusion as a red supergiant, as shown by their large velocities when evolving from, say, 35 to 100~R$_\odot$. 
Such stars are therefore not long-lived as BSGs and not expected to be observed as numerously as other solutions. 
On the other hand, the extended MS stars, blue loop stars, and Case~B binary mergers are all long-lived solutions that may numerously populate this position of the HRD, as evidenced by their small ``velocities'' in the BSG phase. 
Relative to standard single star evolutionary tracks (e.g., those for the classical Hertzsprung gap stars), the extended MS stars are core H-burning stars with oversized cores, and the blue loop and mergers are core He-burning stars with undersized cores. 


\section{Asteroseismology} \label{sec:astero}
We now turn our attention to the asteroseismic properties of these stars, and the prospects of distinguishing the scenarios resulting in BSGs on this basis. 
Small-amplitude, high-overtone stellar oscillations approximately obey the dispersion relation \citep[e.g.,][]{2010aste.book.....A}
\begin{equation}
    \dfrac{\rm{d}^2 \xi_r}{\rm{d}r^2}
    =
    -K \xi_r
\end{equation}
where $r$ is the distance from the centre of the star, $\xi_r$ is the radial displacement wavefunction, and 
\begin{equation} \label{eq:frequencies}
    K
    =
    \dfrac{\omega^2}{c^2}
    \left(
        \dfrac{N^2}{\omega^2} - 1
    \right)
    \left(
        \dfrac{S_\ell^2}{\omega^2} - 1
    \right).
\end{equation}
Here $\omega$ represents the oscillation frequency, $c^2$ is the squared adiabatic sound speed, and $N$ and $S$ are the buoyancy and Lamb frequencies, which are given by
\begin{align}
    N^2
    =
    g
    \left(
        \dfrac{1}{\Gamma_1 p}
        \dfrac{\rm{d}p}{\rm{d}r}
        -
        \dfrac{1}{\rho}
        \dfrac{\rm{d}\rho}{\rm{d}r}
    \right), \qquad
    S^2_\ell 
    = 
    \dfrac{\ell\,(\ell+1)\,c^2}{r^2}
\end{align}
where $g$ is the gravitational acceleration, $\Gamma_1$ the first adiabatic exponent, $p$ the pressure, $\rho$ the density, and $\ell$ the spherical degree of the oscillation mode.
The solutions are oscillatory when
\begin{align}
                   &|\omega|>|N| \;\;\textrm{and}\;\; |\omega|>S_\ell \tag{p modes} \\ 
    \textrm{or}\qquad&|\omega|<|N| \;\;\textrm{and}\;\; |\omega|<S_\ell. \tag{g modes} 
\end{align}
BSGs are observed to pulsate as $\beta$~Cephei variables and $\alpha$~Cygni variables in a superposition of several of both these types of modes \citep[e.g.,][]{2005ApJS..158..193S, 2006ApJ...650.1111S, 2013MNRAS.433.1246S, 2014MNRAS.439L...6G, 2019NatAs...3..760B, 2021ApJ...915..112C, 2021MNRAS.501L..65S, ma_submitted}. 
\mb{Notably, \citet{2006ApJ...650.1111S} detected oscillations in a blue supergiant from observations from the MOST satellite, and \citet{2019NatAs...3..760B} discovered variability in over 100 OB stars with K2 and TESS.}

\citet{1980ApJS...43..469T} studied the stellar oscillation equations asymptotically and found that the typical difference in successive periods $\Delta\Pi$ for high-order $g$ modes (i.e., with overtone $|n| \gg \ell$) is given by
\begin{equation} \label{eq:per-spac}
    \Delta \Pi_\ell \simeq \dfrac{2\pi^2}{\sqrt{\ell\,(\ell+1)}}
    \left(
        \int_{N^2 > 0} \dfrac{N}{r} \; \rm{d}r
    \right)^{-1}.
\end{equation}
Hence, the period spacing $\Delta\Pi_\ell$ for high-order $g$ modes is determined by the shape of the buoyancy cavity, which in turn is particularly sensitive to convective boundaries and molecular weight gradients. 

Similarly, the differences in successive frequencies $\Delta\nu$ for high-order radial $p$ modes is given by 
\begin{equation}
    \Delta\nu \simeq 2 \left(\int_0^R \frac{\rm{d}r}{c} \right)^{-1}
\end{equation}
and thus the $p$-mode frequency separation $\Delta\nu$ depends on the sound travel time across the star, and is well-known to scale with the stellar root mean density. 

\mb{These asteroseismic diagnostics provide routes to determining the evolutionary stage of blue supergiants. 
\citet{2013MNRAS.433.1246S} studied the oscillations of evolutionary models for a similar range of stellar masses as we consider here, and found that both p and g modes are excited by the Fe bump of opacity in MS and post-MS models alike. 
They further found that more radial modes are excited in blue supergiant models in the blue loop phase, offering a potential diagnostic of evolutionary state. We note that mode excitation is an unsolved problem that relies on the considered sources of excitation, the treatment of convection and its interactions with the pulsations, and the assumed opacities and other microphysics.
\mbb{\citet{2019NatAs...3..760B} presented models demonstrating that period spacings can distinguish single stars in the core-hydrogen, shell-hydrogen, and core-helium blue loop evolutionary phases, but did not consider binary models.} } 

Here we compute the oscillation frequencies of our models using the \textsc{Gyre} stellar oscillation code \citep{2013MNRAS.435.3406T, 2018MNRAS.475..879T, 2020ApJ...899..116G, 2023ApJ...945...43S}. 
Propagation diagrams, Kippenhahn diagrams, and individual mode frequencies can be seen in Appendix~\ref{app:grid}. 

\begin{figure*}
    \centering
    \includegraphics[width=0.8\linewidth, trim={0 0.3in 0 0.2in}, clip, keepaspectratio]{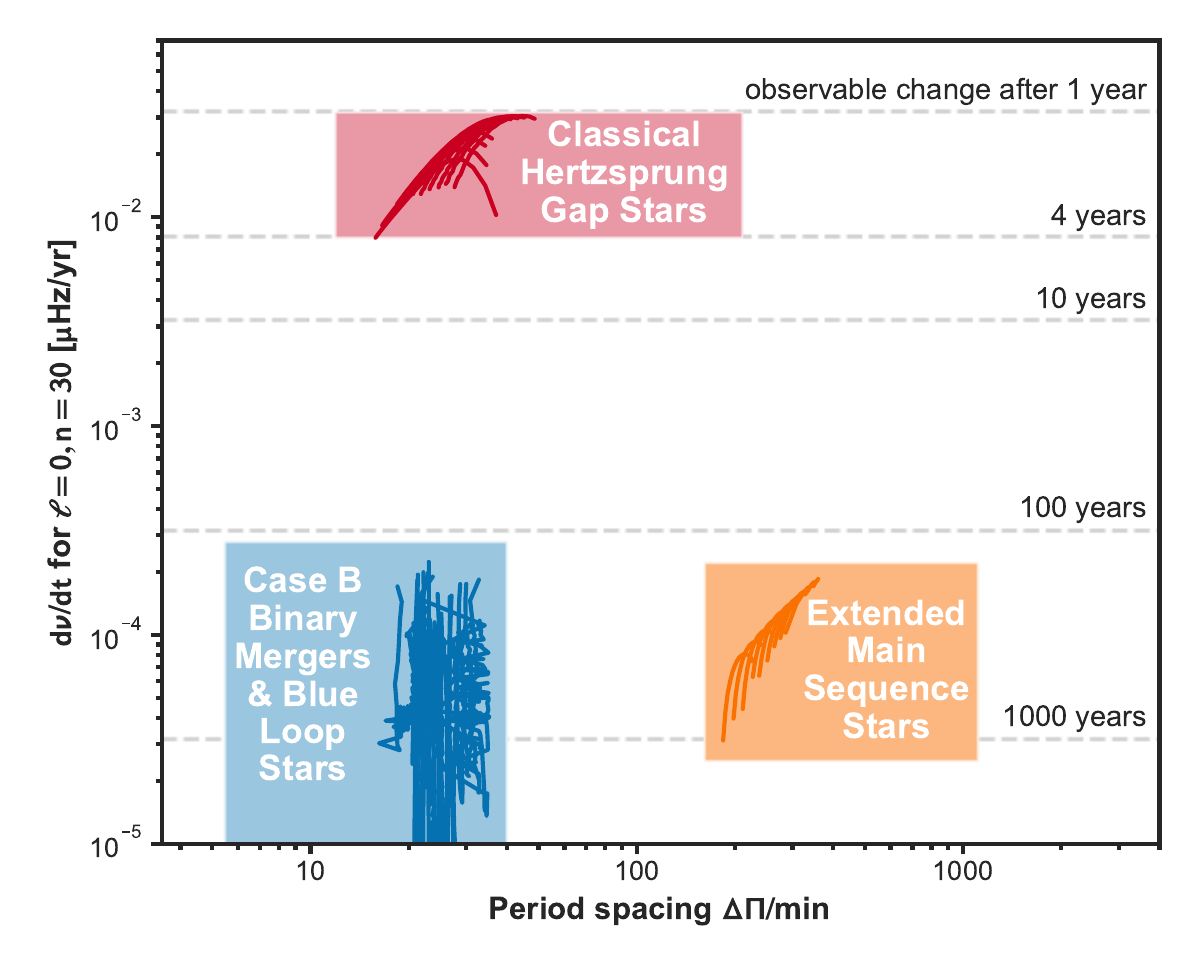}
    \caption{Asteroseismic characteristics of scenarios resulting in blue supergiants. 
    Lines show the evolution of the period spacing and high-order radial oscillation frequencies. 
    The binary mergers models are varied in all input parameters (stellar mass, core mass, mass of the gradient region, and envelope helium abundance).
    Binary mergers and blue loop stars ($\Delta\Pi\sim20$~min) can be asteroseismically distinguished from main-sequence stars ($\Delta\Pi\sim200$~min). 
    These groups can also be distinguished from classical Hertzsprung gap stars ($\rm{d}\nu/\rm{d}t>10^{-2}~\mu$Hz$/$yr) on the basis of high-order radial mode frequency modulation; a star in this mass range with high-order radial mode frequencies that are stable over $>$10~yr cannot be undergoing classical Hertzsprung-gap evolution. 
    \label{fig:radial_change}
    }
\end{figure*}

Figure~\ref{fig:radial_change} shows the evolution of the asymptotic dipolar period spacing $\Delta\Pi\equiv\Delta\Pi_{\ell=1}$, which is a sensitive probe of internal chemical gradients. 
Here it can be seen that stars with H cores stars have $\Delta\Pi\sim200$~min whereas the stars with He cores have $\Delta\Pi\sim20$~min. 
Hence the models predict that the extended main-sequence stars can be clearly distinguished from the other cases on this basis, provided that the period spacing is able to be accurately extracted from observations. 
This analysis is similar to that of \citet{2011Natur.471..608B}, who showed that core-He-burning and shell-H burning low-mass red giants can also be distinguished in a similar fashion. 

An additional feature to distinguish these classes is the rate at which their frequencies change. 
Figure~\ref{fig:radial_change} shows the evolution of high-order radial oscillation frequencies over time. 
While stars in long-lived phases of evolution have stable frequencies, the oscillation modes of classical HG stars decrease in frequency by approximately 0.02~$\mu$Hz/yr. 
Measuring such a change would therefore enable discovery of BSGs in a rapid evolutionary phase and thus opens the possibility of finding genuine classical HG stars. 
This is similar to the analysis of \citet{1920Obs....43..341E}, who showed that the stability of Cepheid periods over time is incompatible with the contraction theory of stellar evolution. 

Frequency modulation becomes measurable when the change is greater than the frequency resolution, which is given by the reciprocal of the total observing time $1/T$. 
For a 5$\sigma$ detection with instruments such as the ongoing NASA TESS mission \citep{2015JATIS...1a4003R} it would be possible to identify some classical HG stars from 5~yr of observations. 
It will furthermore be possible to characterize the entire population from 20~yr of observations, which may eventually be possible with TESS, or otherwise possible in combination with historical data. 

Taken together, these two diagnostics imply that it is theoretically possible to distinguish three of the cases on a star-by-star basis using asteroseismology. 
It must be noted that carrying out this analysis observationally may pose challenges. 
The relevant oscillation modes must have sufficient amplitude as to be observable. 
As we carry out linear pulsation calculations, we are not able to predict the amplitudes of the oscillations. 
Furthermore, extracting period spacings from observations may require accurate mode identification, which is often a challenge in massive star asteroseismology. 

\needspace{3\baselineskip}
\section{Discussion} \label{sec:discussion}

Here we have investigated four scenarios that can result in blue supergiant stars:
\begin{enumerate}
    \item \textbf{Classical Hertzsprung gap stars}: Post-main sequence stars that cross the HG on a thermal timescale and thus are not expected to numerously populate this position of the HRD (see Figure~\ref{fig:hrd}). 
    
    \item \textbf{Extended main-sequence stars}: Massive central H-burning stars with oversized cores. They may be the result of very significant extra mixing beyond the convective core, arising for example from rotation or a prior binary interaction. 
    
    \item \textbf{Blue loop stars}: Stars with He-burning cores who spend a significant fraction of their post-MS evolution as a BSG. These are found in models of single star evolution that assume little or no overshooting and very efficient semi-convection. 
    
    \item \textbf{Case B binary mergers}: A post-MS star merging with a MS star can similarly lead to a long-lived BSG with a He-burning core and a H-burning shell. A similar structure may also result from accretion onto a evolved star. 
\end{enumerate}
Observations have revealed an overdensity of BSGs in the HG, indicating that a long-lived solution is required, and thus ruling out classical HG stars \mb{as their primary channel of origin}. 
The extended main-sequence star scenario is difficult to understand from a standpoint of single star evolution, as the requisite convective core overshoot appears to be much greater than is supported by observations \citep{2011A&A...530A.115B, 2021A&A...655A..29J, 2023NatAs...7..913B}. 
Even then it can only explain the hottest BSGs. 
However, as noted above, stars with similar structures may be created in binary systems, such as through Case~A mergers. 
The blue loop solution has the drawback that it fails to predict the width of the MS, and also that it predicts an additional gap between the MS and the blue supergiants (see Figure~\ref{fig:hrd}) that is not observed. 

Although stellar mergers seem like they would be very rare, observations have revealed that massive stars regularly form with close companions, and as many as one in three eventually merge \citep{2012Sci...337..444S, 2014ApJ...782....7D, 2019A&A...631A...5Z}. 
\mb{Recently, \citet{2023arXiv231112124H} computed a grid of binary evolution models and presented numerous pathways to such mergers.} 
Case~B binary mergers therefore constitute a highly plausible explanation for the overdensity of BSGs in the HG. 

In this work, we presented grids of stellar evolution and pulsation models for the above scenarios with the aim of determining whether it is possible to separate these cases observationally. 
We found two asteroseismic diagnostics: the period spacing and the frequency modulation over time, which appear capable of cleanly separating classical HG stars and MS stars from the others (see Figure~\ref{fig:radial_change}). 

Distinguishing Case~B mergers from blue loop stars remains a challenge. 
As these scenarios result in stars with a similar structure, they are difficult to separate based on their global asteroseismic quantities such as $\Delta\Pi$ (see also the propagation and Kippenhahn diagrams in Appendix~\ref{app:grid}). 
Since blue loop models are similarly long-lived in this phase as merger models, the frequency modulation signal is similarly small. 
A detailed analysis of individual mode frequencies would therefore be necessary in order to determine the detailed differences in internal structure between these two scenarios. 
Such an analysis may be possible to carry out through an inverse analysis of stellar oscillation data \citep{2017ApJ...851...80B, 2019ApJ...885..143B, 2021ApJ...915..100B}; the development of these techniques for massive pulsators is underway \citep{2023arXiv230509624V}. 
It may also be possible to distinguish these cases on the basis of their surface chemical compositions \citep{2013MNRAS.433.1246S, 2014MNRAS.439L...6G}. 
\mb{Recently, \citet{2024ApJ...963L..42M} compared the compositions of merger models with a sample of BSGs in the LMC and found that the mergers are better able to reproduce the nitrogen, carbon, and oxygen abundance ratios than single star models.} 

A companion paper to this one investigates the variability of BSGs in the Large Magellanic Cloud observed by the TESS spacecraft \citep{ma_submitted}, making use of the sample presented by \citet{Serebriakova2023} \mb{following the discovery papers of \citet{2019NatAs...3..760B, 2019A&A...621A.135B}}. 
They find ubiquitous signals of stochastic low-frequency variability in BSG stars, which may arise from the $g$-mode pulsations studied in this work. 
Follow-up work is required however to further characterize these signals and to understand whether period spacings can be measured from them.

The work presented here is among a now emerging literature using asteroseismology to investigate binaries and binary merger products. 
\citealt{2021MNRAS.508.1618R} investigated the possibility of identifying stellar merger remnants through asteroseismology of low-mass red-giant stars. They demonstrated that low-mass Case~B merger remnants can be distinguished from single stars by measuring the dipolar period spacing. 
Our work confirms this behavior in higher-mass stars. 

\citealt{2022NatAs...6..673L} used asteroseismology to identify two classes of stars among $\sim$7,000 helium-burning red giants observed by NASA's Kepler mission that must have undergone significant mass loss, presumably due to stripping in binary interactions. 
\citealt{2022A&A...659A.106D} looked at the seismic signature of electron degeneracy in the cores of red giants, providing insights into the evolution of red giants resulting from interaction between two low-mass close companions during the red giant branch phase. 
Mergers have also long been invoked to explain SX~Phoenicis stars \citep{1990ASPC...11...64N} which are blue straggler stars, i.e., stars that pulsate as if they are young stars while belonging to old globular clusters. 

Those investigations mainly focus on low-mass stars, largely because these were the stars with the best observations coming out of the NASA \emph{Kepler} mission \citep{2010Sci...327..977B}. 
In contrast, we focus here on the evolution of massive stars. 
Since \emph{Kepler} pointed at only a small field of view, and massive stars are relatively rare, few massive stars have been characterized asteroseismically. 
The full-sky coverage from TESS has resulted in more observations of massive stars, albeit with larger pixels and degraded photometric precision. 
As of this writing, TESS has been observing for $\sim$5~years, and so the effects discussed herein are in principle soon becoming observable. 
The upcoming PLATO mission \citep{2017AN....338..644M} will offer an even greater opportunity to characterize blue supergiants due to its increased photometric precision and large field of view. 

A potential limitation of our analysis is that we consider only non-rotating stellar models and non-rotating solutions to the stellar oscillation equations. 
This may be justified, however, as BSGs are typically very slow rotators \citep[e.g.,][]{2008A&A...479..541H}, and stellar mergers seem to result in slowly-rotating stars \citep[e.g.,][]{2019Natur.574..211S}. 
\mb{We note however that even relatively slow rotation can cause the period spacing to become a negatively sloped function of the pulsation period \citep{2020FrASS...7...70B}, which needs to be taken into account when modeling observed stars.} 
Also, our analysis of the observability of the radial mode frequency change over time assumed that the modes are coherent. 
Modes such as those seen in the Sun are stochastically excited and damped; the finite lifetime of these modes leads to an intrinsic mode width in a periodogram, which would complicate a frequency modulation analysis. 
It is conceivable that other effects, such as a magnetic cycle, could induce frequency changes on a multi-year timescale; such effects could possibly be disentangled via spectropolarimetry and other techniques. 
Detailed modelling of observed stellar oscillation frequencies will therefore be necessary to account for all possible effects when trying to determine the evolutionary state of a blue supergiant.

\section{Conclusions} \label{sec:conclusion}
The nature of blue supergiants remains an enigmatic unsolved problem in astrophysics. 
As they are among the most luminous stars in the Universe, this is a problem that is not only of concern to stellar theorists but one that affects many fields in astrophysics that directly or indirectly rely on evolutionary tracks of massive stars. 

The large occurrence rate of close binaries among massive stars, together with the much longer lifetimes of binary mergers than single stars passing through the BSG phase, strongly suggests that a very significant fraction, or even the majority, of them may be binary merger products. Solving the BSG problem will thus almost certainly require understanding and distinguishing the products of single and binary evolution. 

In this paper, we explore several scenarios for BSGs and explore whether asteroseismology may help to to distinguish them. Our takeway points are:
\begin{enumerate}
    \item \textbf {Explaining long-lived blue supergiants appears to require stars with abnormal interior structures as may result from binary interaction.} 
    Consistent with earlier work, our models illustrate (see Figure~\ref{fig:hrd}) that long-lived solutions for blue supergiant stars arise either from \textbf{core-H-burning stars with oversized cores} (as may result from enhanced overshooting, Case~A mergers, or accretion onto unevolved mass gainers) or \textbf{core-He-burning stars with undersized cores} (as may result from supressed overshooting, Case B mergers, or accretion onto evolved mass gainers).    
    
    \item \textbf{Asteroseismic diagnostics have the potential to distinguish between different scenarios for BSGs; we expect almost an order of magnitude difference in the period spacing for $\textbf{g}$-modes for these two main scenarios.} 
    For stars with H cores (e.g., extended main sequence stars) we predict ${\Delta\Pi \sim 200}$~min whereas for stars with He cores (classical HG stars, binary mergers, and blue loop stars) we predict ${\Delta\Pi \sim 20}$~min (see Figure~\ref{fig:radial_change}). 
    This very large difference appears robust against model variations (see Figure~\ref{fig:astero}).
    
    \item \textbf{We suggest searching for frequency changes in high-radial-order modes as a signature of thermal timescale evolution.} For BSG stars expanding raipdly during H-shell burning (classic Hertzsprung gap stars) we predict the changes in the high-radial-order modes to be of the order of $10^{-2}~\mu$Hz/yr, which is potentially detectable (see Figure~\ref{fig:radial_change}).  
\end{enumerate}

Although our models for mergers are schematic at best, our findings seem robust against the wide range of model variations that we explored. This gives confidence that we have identified a solid diagnostic signature which can be probed by asteroseismology. 

Whether in practice the modes can be observed and identified remains to be seen. It will require prioritizing long timeseries observations of BSGs. Multi-color photometry or contemporaneous spectroscopic timeseries would aid in mode identification. 

It is very promising that TESS is currently observing numerous blue supergiants, and will continue doing so for the foreseeable future \citep{2019NatAs...3..760B, 2020FrASS...7...70B, ma_submitted}. 
The PLATO mission \citep{2014ExA....38..249R}, set to launch in 2026 as an all-sky successor to the \emph{Kepler} mission, has the potential to further revolutionize our understanding of massive star evolution, and hopefully solve the blue supergiant problem.

\acknowledgements The authors thank the anonymous referee for their helpful suggestions that have improved the manuscript. The authors are grateful to Conny Aerts, Max Briel, Abel de Burgos Sierra, Jim Fuller, Evert Glebbeek, Natasha Ivanova, Cole Johnston, Linhao Ma, Ehsan Moravveji, Sergio Sim\'on-D\'ias, Philipp Podsiadlowski, Mathieu Renzo, and Chen Wang for helpful discussions. 
\software{} Python, Jupyter, MESA, GYRE%

\appendix

\section{Merger Model Grid} \label{app:grid}
\begin{figure*}[hb]%
    \centering%
    \includegraphics[width=\linewidth, trim={0 9.2cm 0 0}, clip, keepaspectratio]{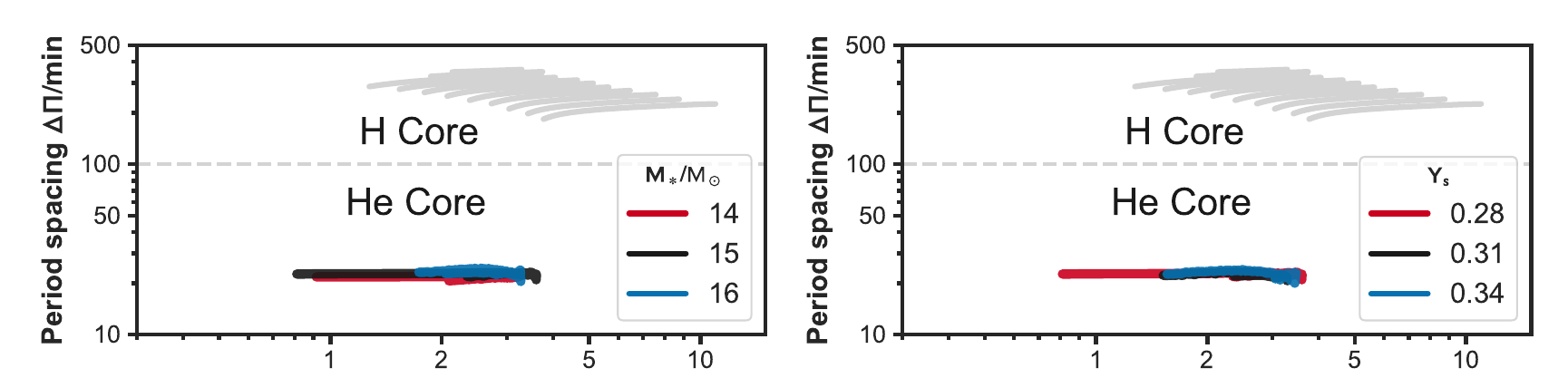}\\%
    \includegraphics[width=\linewidth, trim={0 0 0 8cm}, clip, keepaspectratio]{figs/astero.pdf}%
    \caption{Asteroseismic diagnostics for the two main classes of long-lived blue supergiant solutions: H-burning and He-burning stars, shown in the asteroseismic $\Delta\nu$--$\Delta\Pi$ plane. The large frequency separation $\Delta\nu$ is sensitive to the mean density of the star, and the period spacing $\Delta\Pi$ probes internal chemical composition gradients. 
    The figure is split into four panels representing the four free parameters of the binary merger model: stellar mass (upper left), envelope helium abundance (upper right), mass of the gradient region (lower left), and core mass (lower right; see Section~\ref{sec:merger} for definitions). 
    Across all variations in input parameters, binary mergers have $\Delta\Pi<50$~min and H-core burning stars have $\Delta\Pi>100$~min.
    Only long-lived phases of evolution are shown. 
    Evolution generally proceeds from high to low $\Delta\nu$ (i.e., right to left), and more massive stars have larger period spacings. 
    \label{fig:astero} }%
\end{figure*}

In Figure~\ref{fig:astero} we show the evolution of binary merger products in the diagnostic $\Delta\nu$--$\Delta\Pi$ diagram under variations in the initial model parameters. 
Both of these quantities are potential observables, provided that the modes are excited to observable amplitudes and can be identified. 
Note that binary mergers remain distinguishable from H-core stars on the basis of the period spacing across changes to the input parameters of the merger model. 

\begin{figure}
    \centering
    \includegraphics[width=0.5\linewidth]{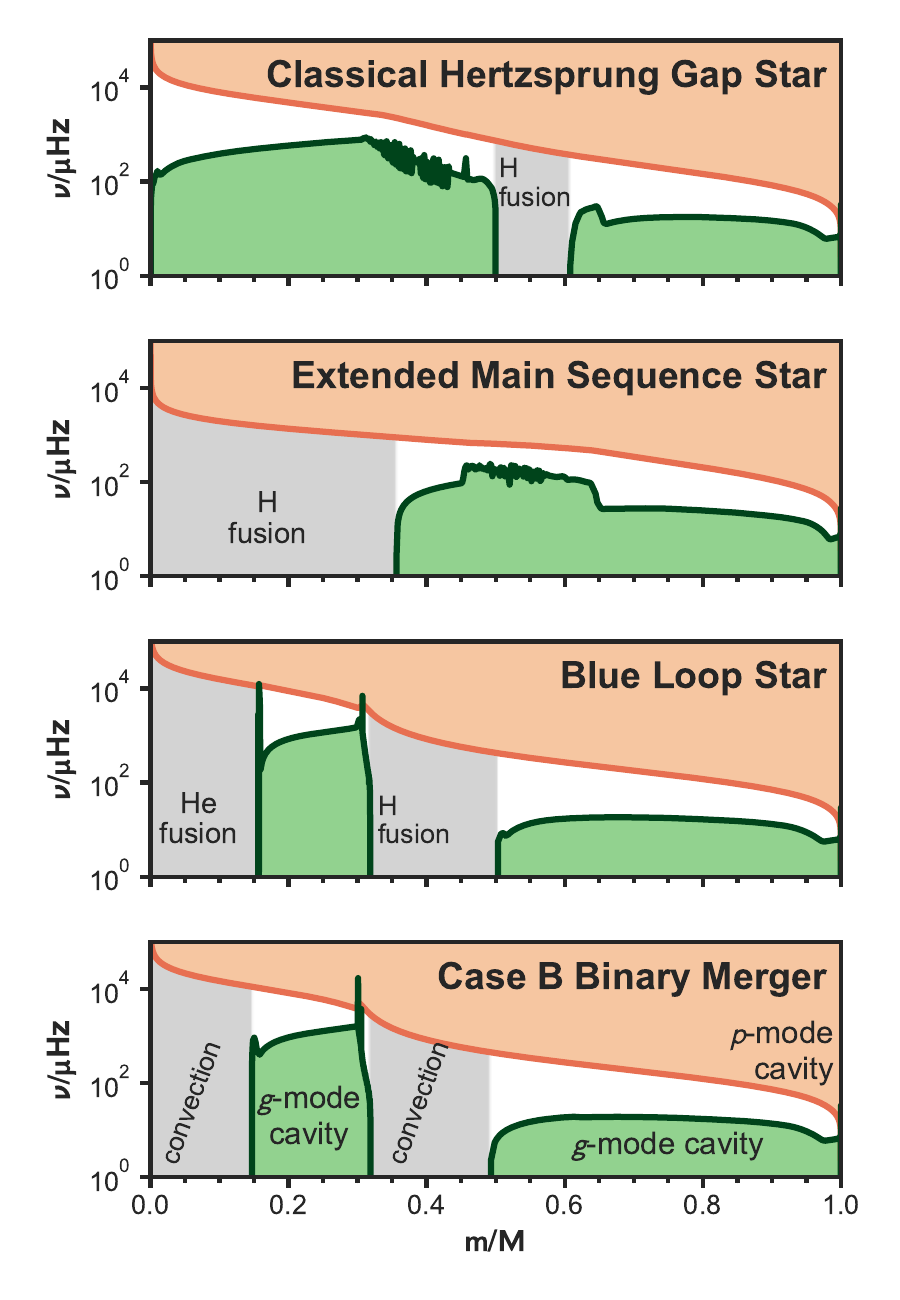}%
    \includegraphics[width=0.5\linewidth]{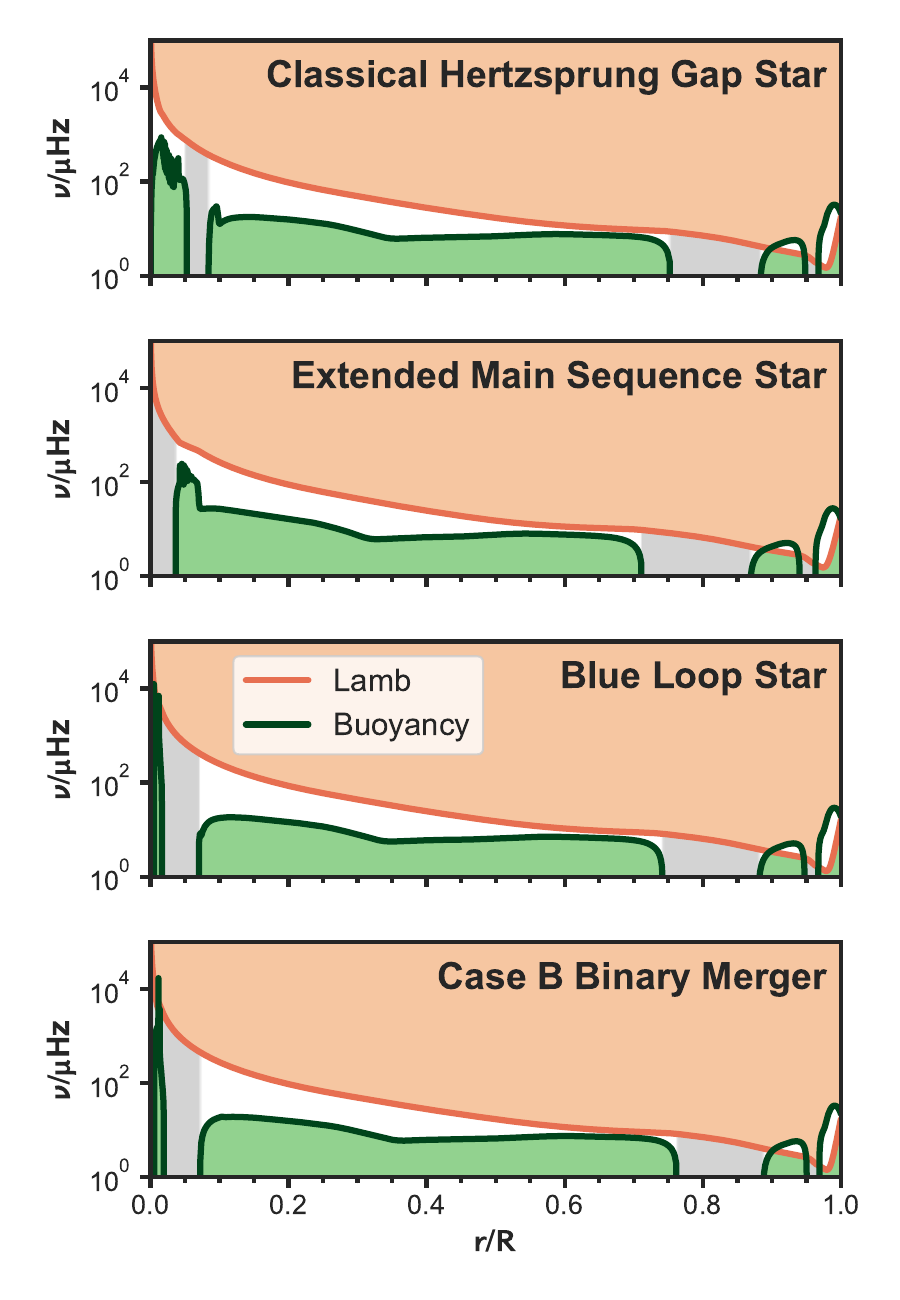}%
    \caption{Propagation diagram showing the buoyancy ($N$) and Lamb ($S_{\ell=1}$) frequency profiles (Equation~\ref{eq:frequencies}), which determine the asteroseismic characteristics, as a function of fractional mass (left panels) and fractional radius (right panels) for stellar models of ${17~\rm{M}_\odot}$ and ${35~\rm{R}_\odot}$. 
    The buoyancy frequency is imaginary in convective zones and hence shows, for example in the merger model: the convective He-burning core, intermediate convection zone above the H-burning shell, and two subsurface convection zones. 
    The overlap of the $p$ and $g$ mode cavities in the near-surface layers causes most of the dipole modes to be mixed modes. 
    The extended MS stars have higher $\Delta\Pi$ (here 170~min, rather than 30~min for the other models) because of their larger $\int N/r~\textrm{d}r$. } 
    \label{fig:propagation}
\end{figure}

\begin{figure*}[ht]%
    \centering%
    \includegraphics[height=10cm, trim={0 0 3.3cm 0}, clip, keepaspectratio]{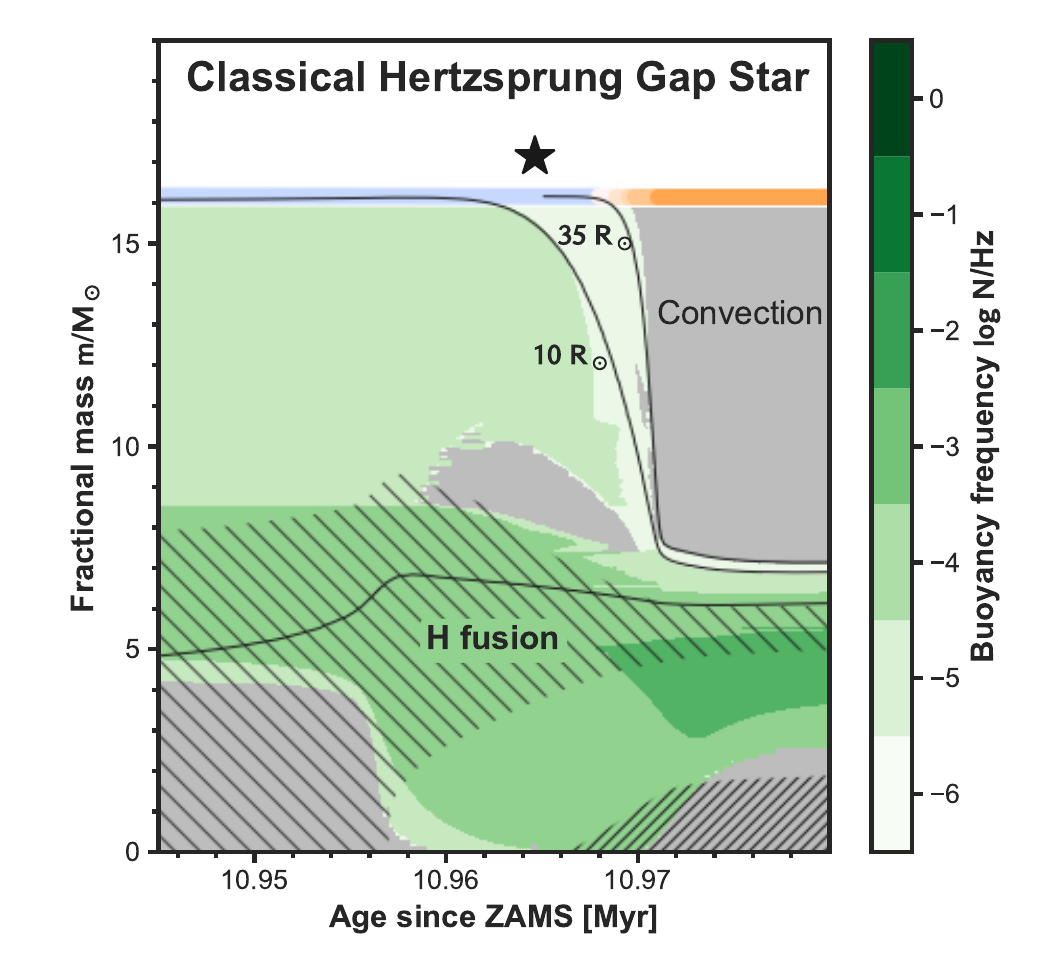}%
    \includegraphics[height=10cm, trim={2.4cm 0 0 0}, clip, keepaspectratio]{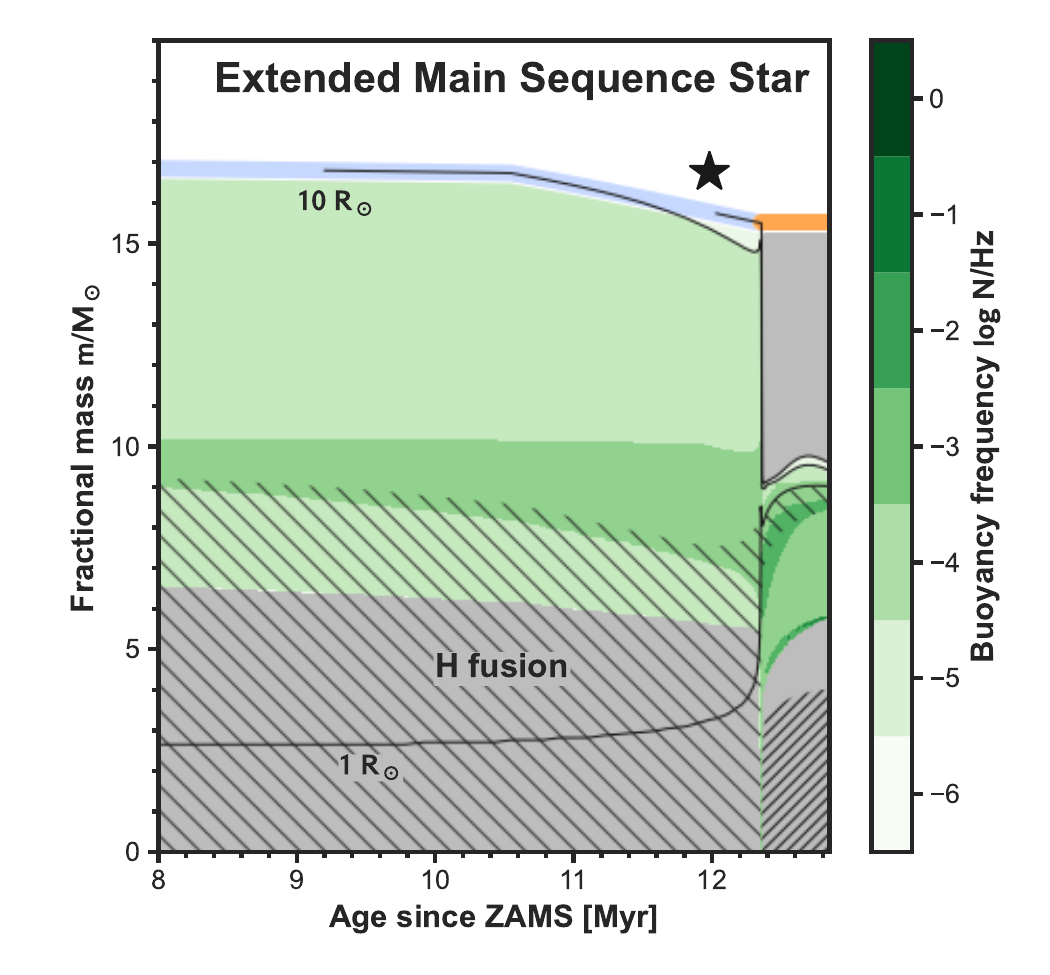}\\%
    \includegraphics[height=10cm, trim={0 0 3.3cm 0}, clip, keepaspectratio]{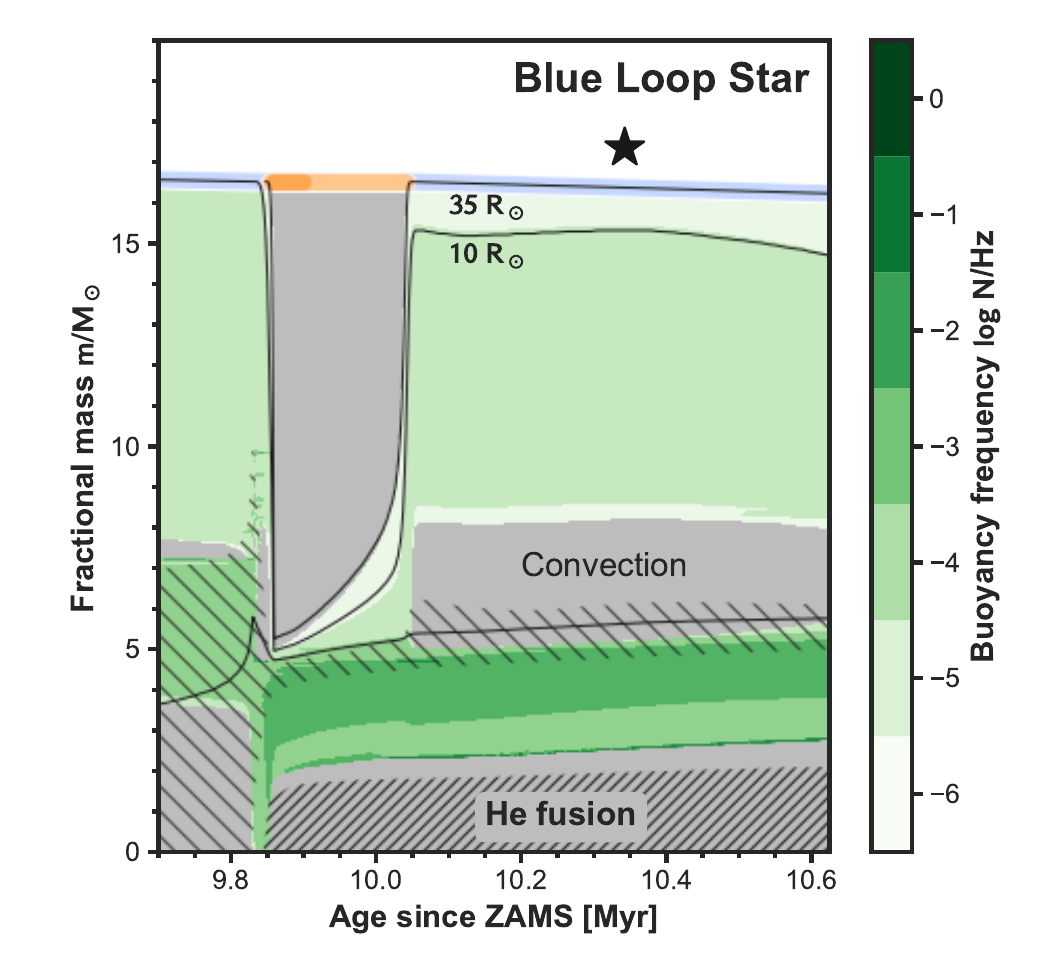}%
    \includegraphics[height=10cm, trim={2.4cm 0 0 0}, clip, keepaspectratio]{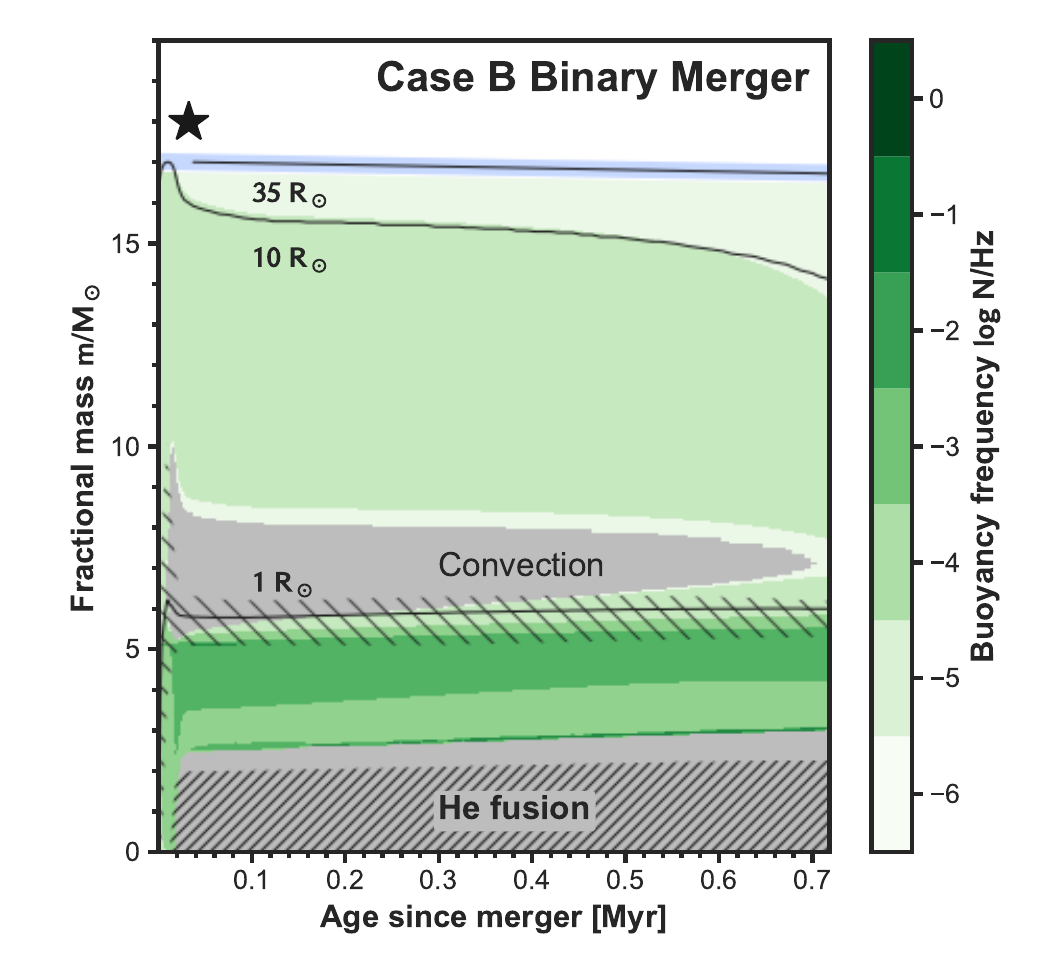}%
    \caption{Kippenhahn diagrams showing the evolution of stars through the blue supergiant phase, all with an initial mass of 17~M$_\odot$. 
    The classical HG star is short-lived (0.01 Myr, upper left panel) while the other solutions are all long-lived. 
    Convective regions are shown in gray; g-modes are exponentially damped in these regions. 
    Regions of nuclear fusion are indicated by slashes. 
    The buoyancy frequency is given in green; a sharp peak is visible below the H-burning shell of the merger model. 
    The photosphere of each model is colored by the spectral type. 
    Contours of constant radius are shown for 1, 10, 35~R$_\odot$, the latter of which is the approximate radius of a BSG in this mass range. 
    The black star above the photosphere in each diagram indicates which blue supergiant model is further visualized in Figure~\ref{fig:propagation}. 
    For an example of how to read these diagrams, the classical HG star is shown passing from having an H-burning core into a brief phase as a BSG with a H-burning shell for $\sim10$~kyr, followed by entering the RSG phase with a large convective envelope, a He-burning core, and an H-burning shell. 
    \label{fig:kippenhahn} }%
\end{figure*}

\begin{figure*}[ht]%
    \centering%
    \includegraphics[height=10cm, trim={0 0 3.3cm 0}, clip, keepaspectratio]{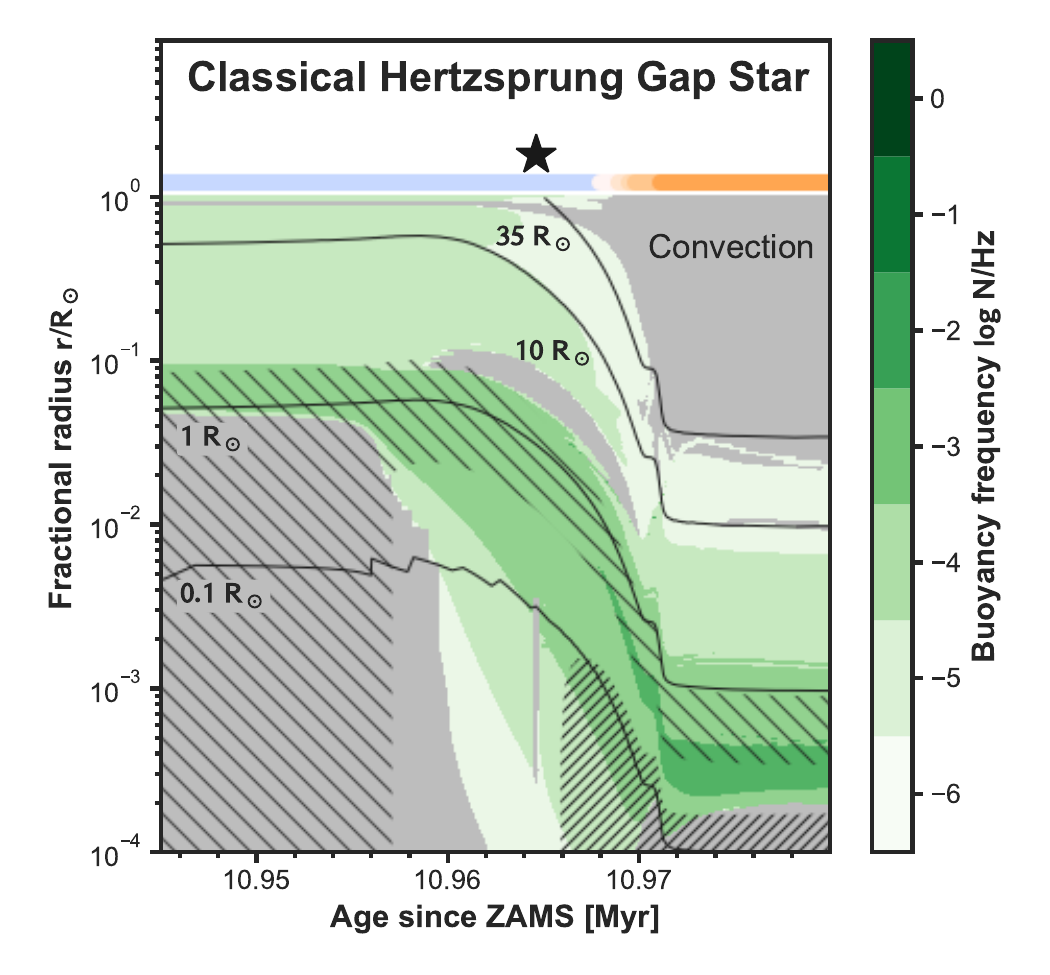}%
    \includegraphics[height=10cm, trim={2.4cm 0 0 0}, clip, keepaspectratio]{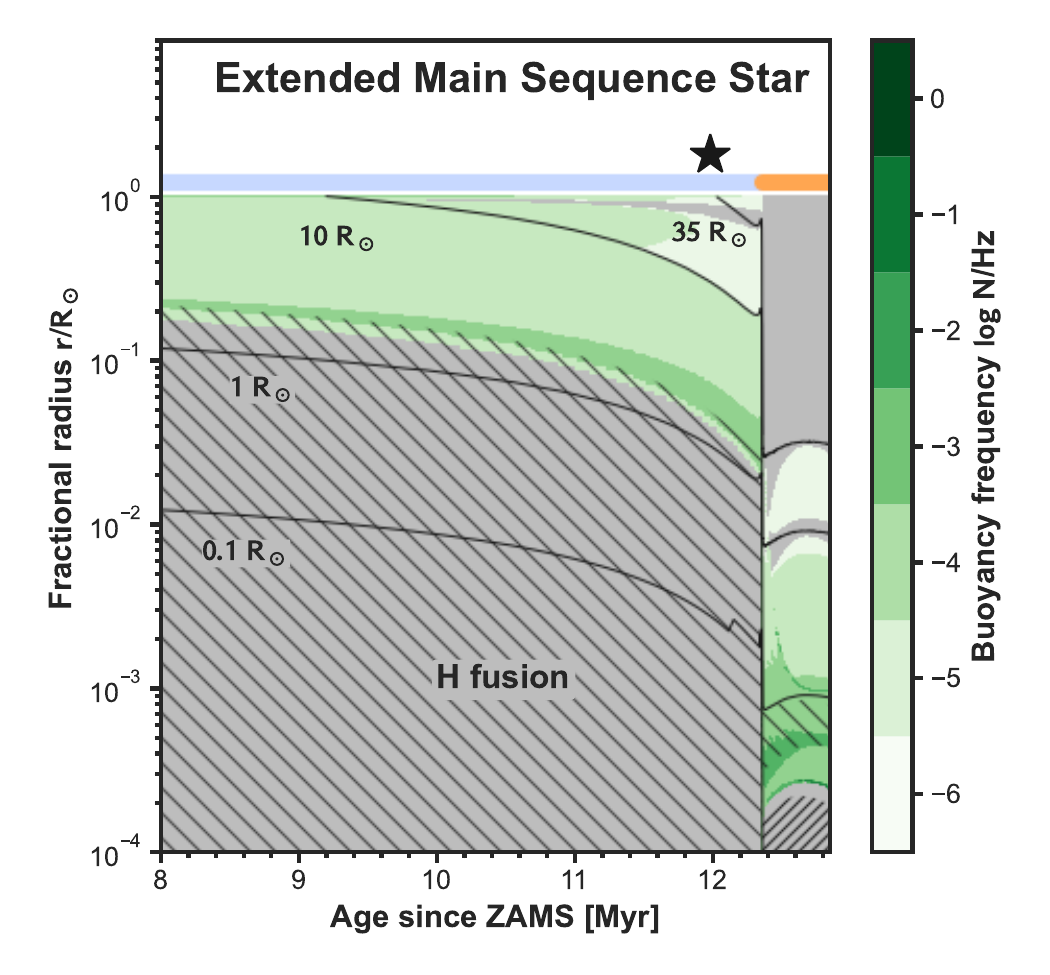}\\%
    \includegraphics[height=10cm, trim={0 0 3.3cm 0}, clip, keepaspectratio]{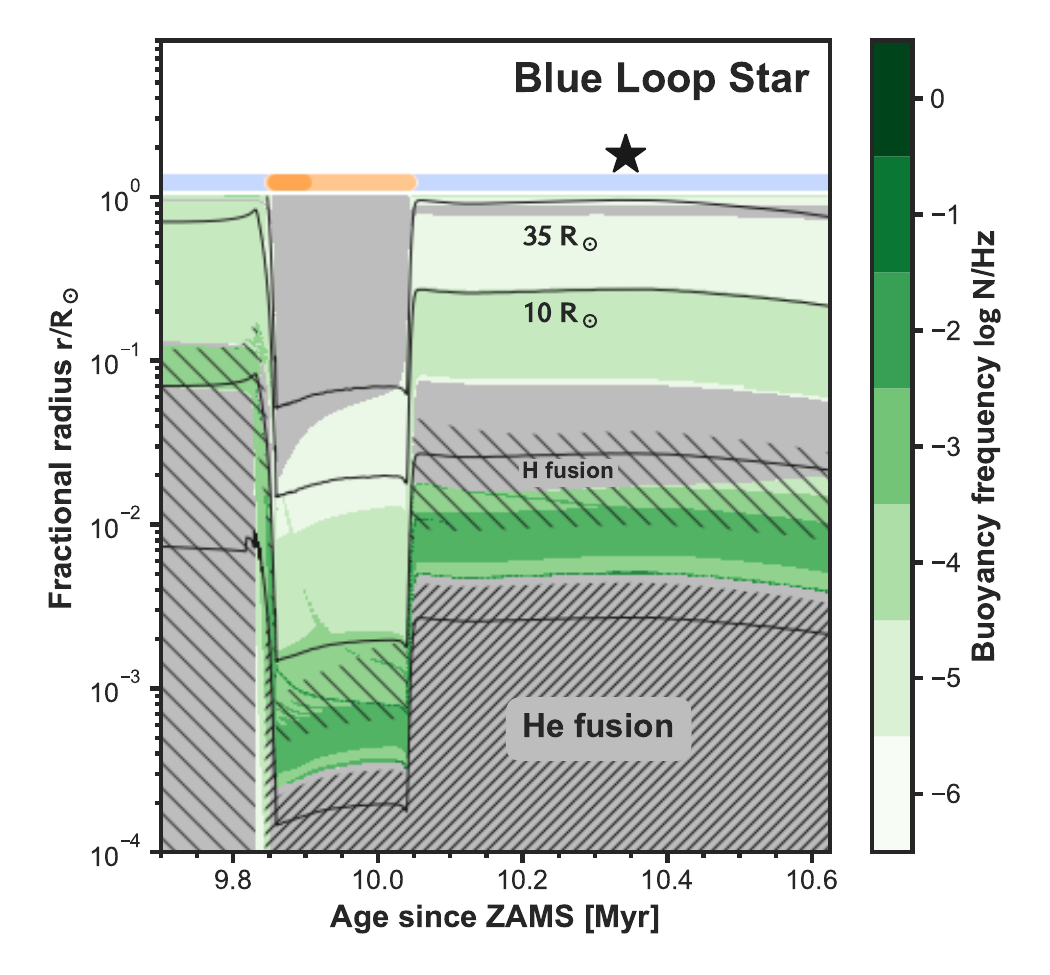}%
    \includegraphics[height=10cm, trim={2.4cm 0 0cm 0}, clip, keepaspectratio]{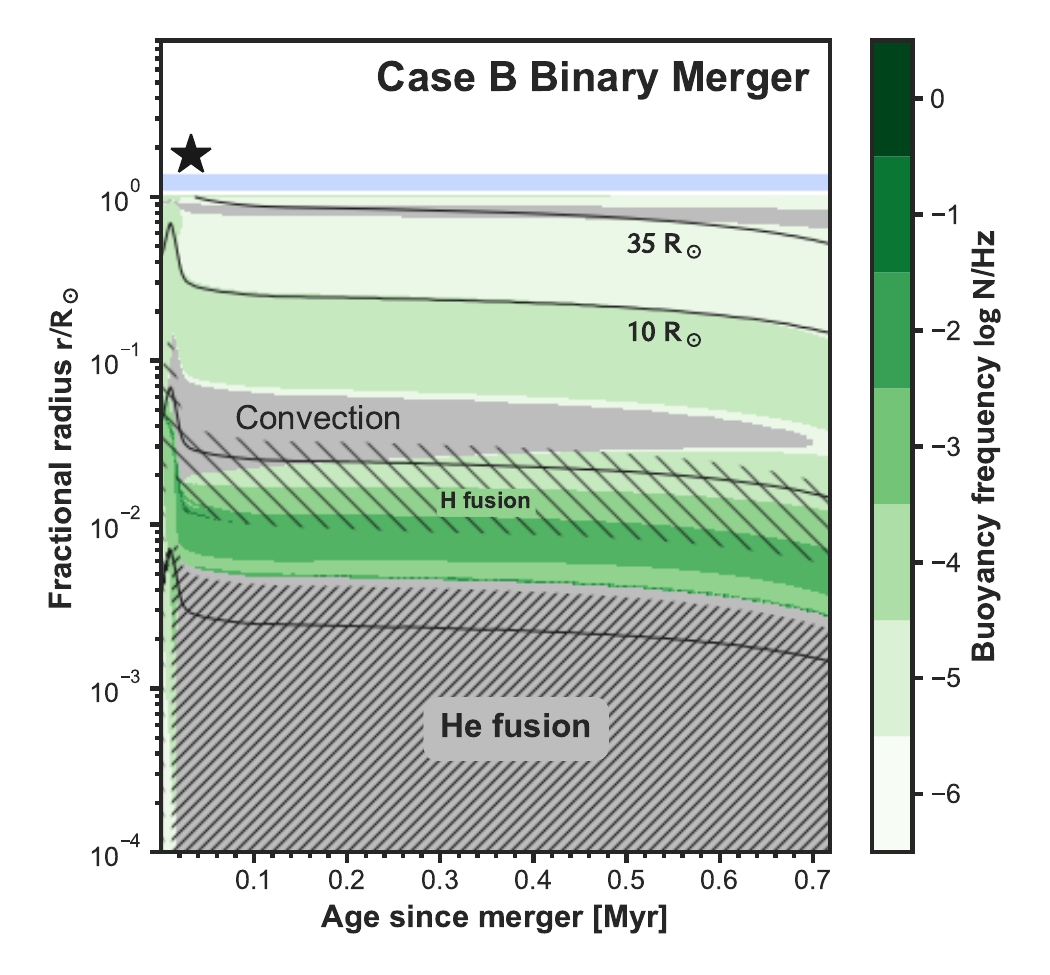}%
    \caption{The same as Figure~\ref{fig:kippenhahn}, now shown in terms of fractional radius. Here contours of constant radius are also shown for 0.1~R$_\odot$. 
    \label{fig:kippenhahn_r} }%
\end{figure*}

\begin{figure}
    \centering
    \includegraphics[width=0.5\textwidth, trim={0 2cm 0 0.5cm}, clip]{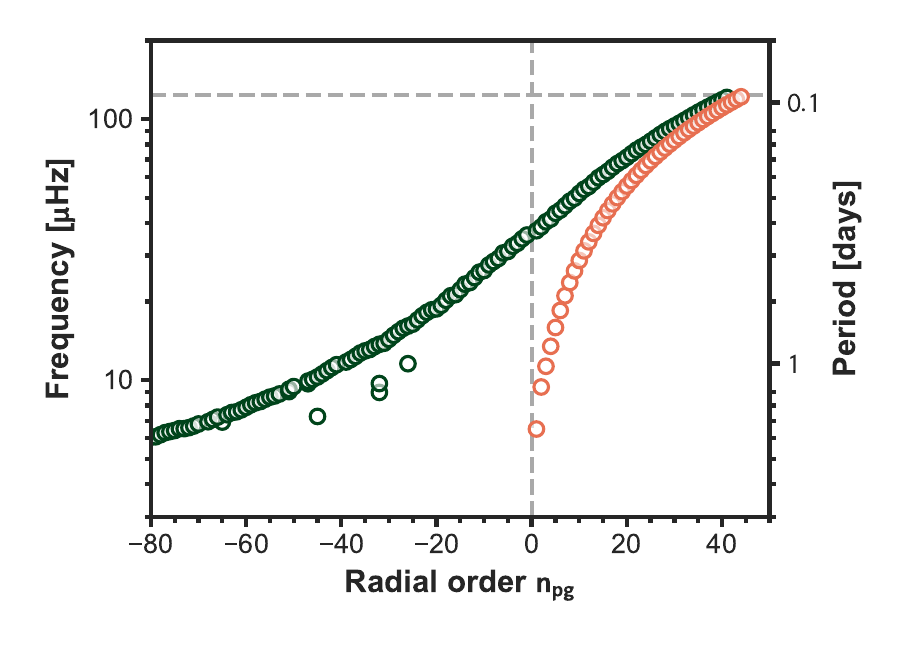}%
    \includegraphics[width=0.5\textwidth, trim={0 2cm 0 0.5cm}, clip]{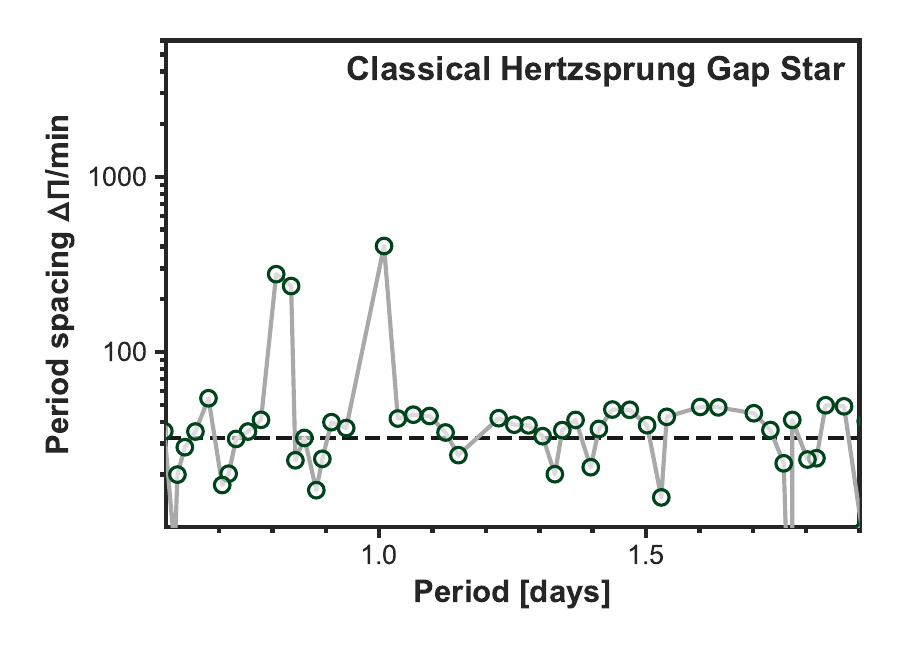}\\%
    \includegraphics[width=0.5\textwidth, trim={0 2cm 0 0.5cm}, clip]{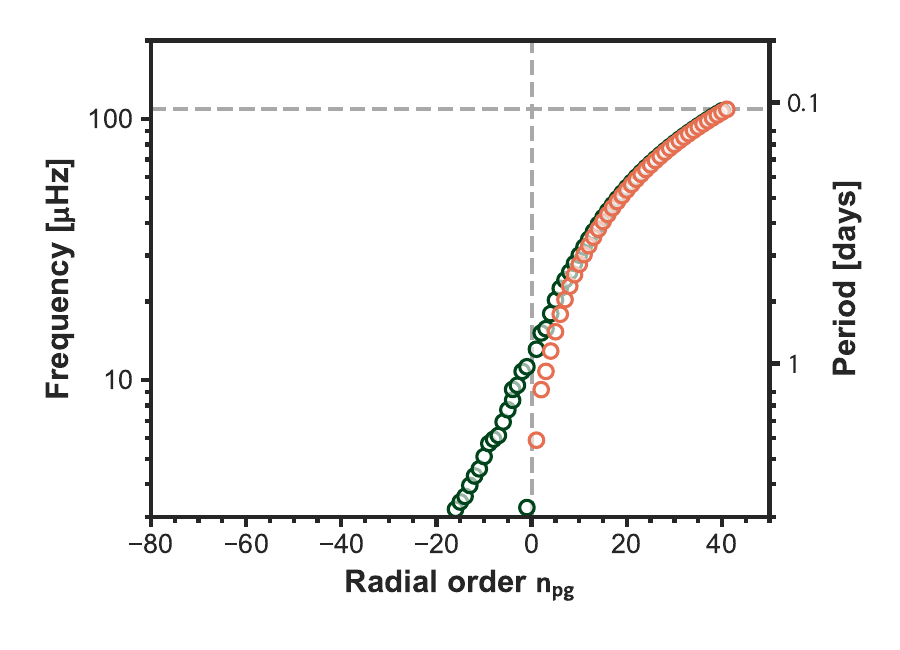}%
    \includegraphics[width=0.5\textwidth, trim={0 2cm 0 0.5cm}, clip]{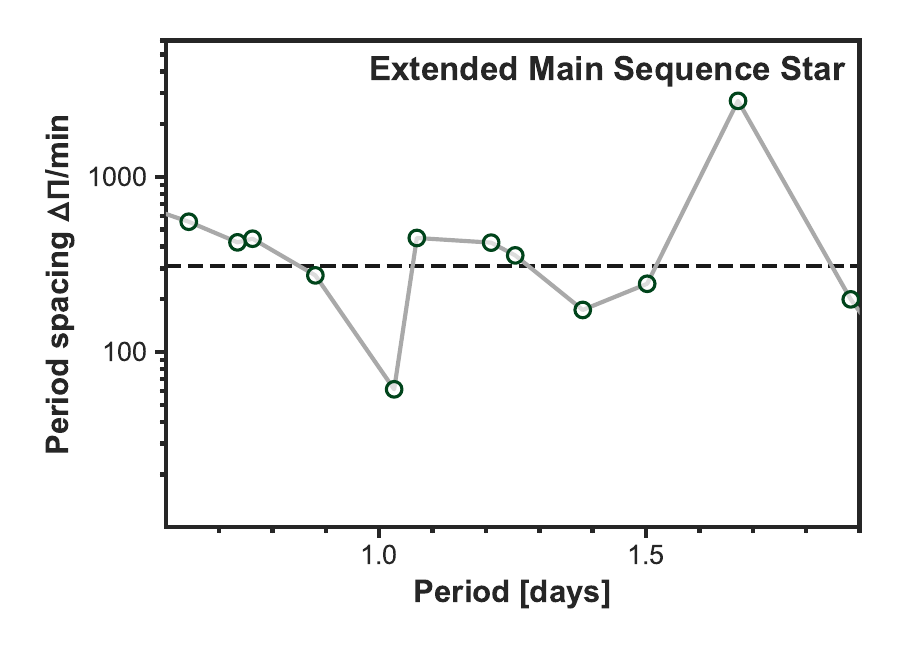}\\%
    \includegraphics[width=0.5\textwidth, trim={0 2cm 0 0.5cm}, clip]{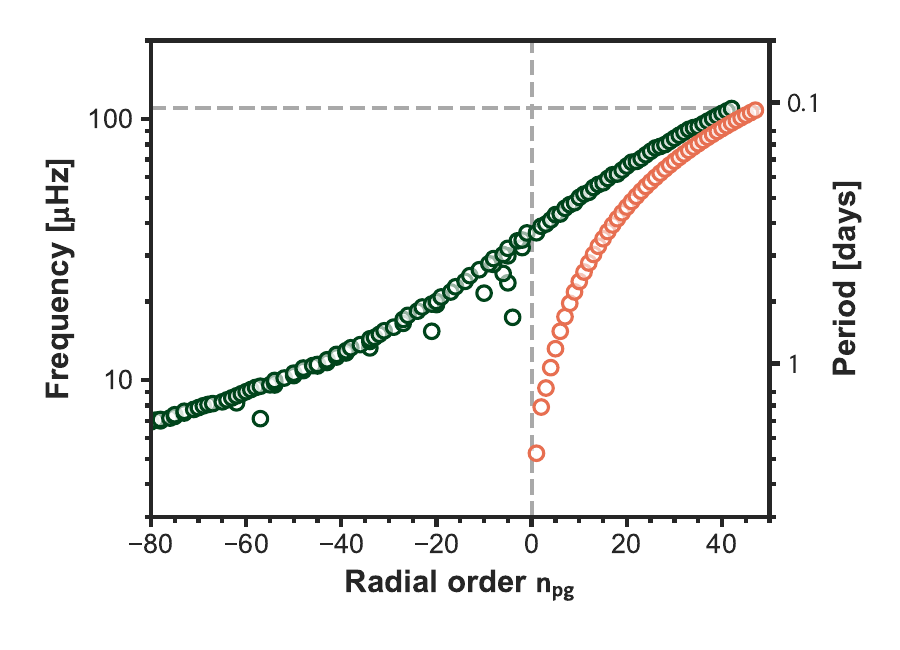}%
    \includegraphics[width=0.5\textwidth, trim={0 2cm 0 0.5cm}, clip]{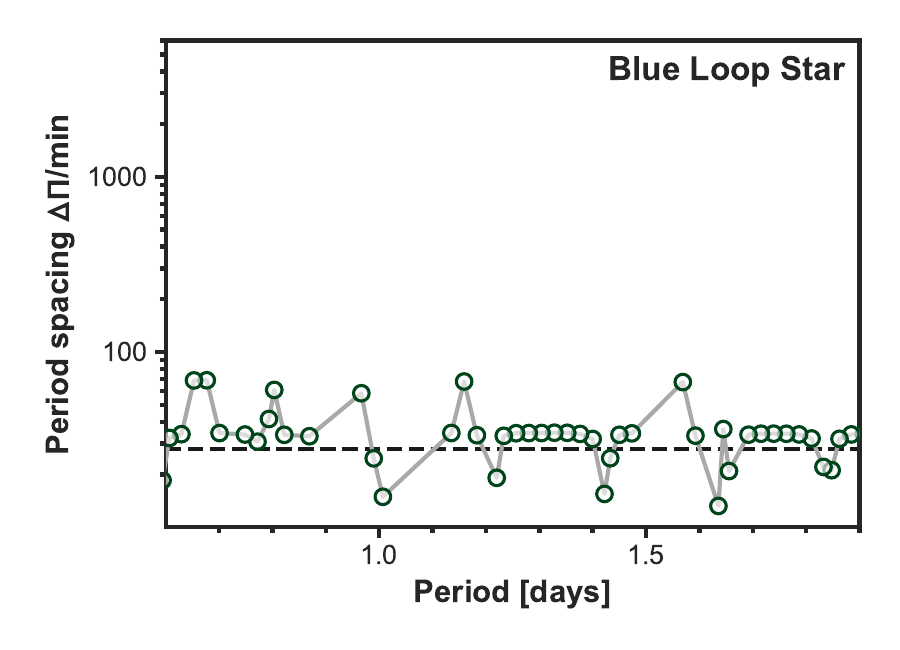}\\%
    \includegraphics[width=0.5\textwidth, trim={0 0 0 0.5cm}, clip]{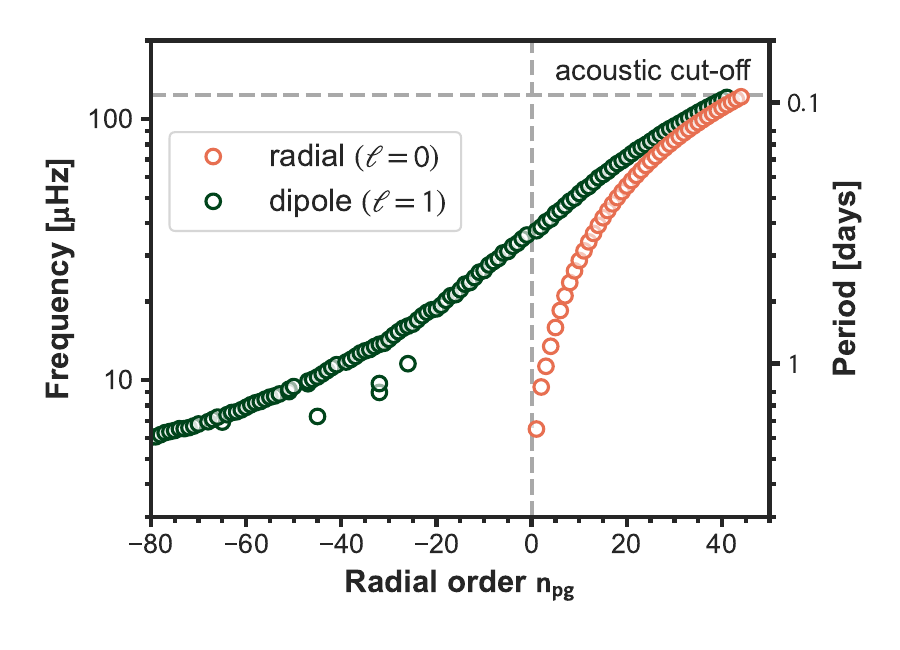}%
    \includegraphics[width=0.5\textwidth, trim={0 0 0 0.5cm}, clip]{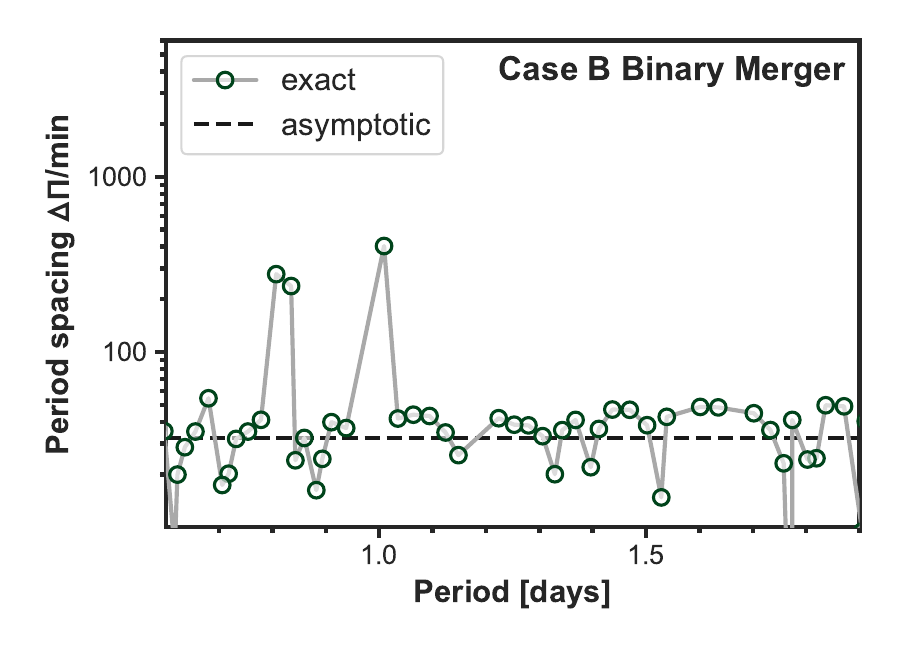}\\%
    \caption{\textsc{Left Panel}. Radial and non-radial oscillation mode frequencies for the blue supergiant models shown in Figure~\ref{fig:propagation}, plotted as a function of the radial order. By convention $g$-dominated modes have $n<0$. 
    \textsc{Right Panel}. Period spacings for the $g$ mode frequencies of these models from \textsc{Gyre}, shown in comparison with the asymptotic value. 
    Note that most of the dipole modes are mixed modes. 
    The extended MS model has a period spacing of $\sim$300~min while the others have period spacings of $\sim$30~min. 
    \label{fig:frequencies}}
\end{figure}

The evolution of the period spacing for blue supergiants can be further understood by examining a Kippenhahn diagram of the internal buoyancy frequency $N$ (see Figures~\ref{fig:propagation}, \ref{fig:kippenhahn}, and \ref{fig:kippenhahn_r}). 
Since $N$ remains fairly constant throughout the evolution of the merger, so too does the period spacing. 
Finally, the individual oscillation mode frequencies and individual period spacings for the models shown in Figure~\ref{fig:propagation} are visualized in Figure~\ref{fig:frequencies}. 

\clearpage
\bibliographystyle{aasjournal.bst}%
\bibliography{main}%

\end{document}